\documentclass[%
aps,
 amsmath,amssymb,
reprint,%
]{revtex4-1}

\usepackage{graphicx}
\usepackage{dcolumn}
\usepackage{bm}
\usepackage{ulem}
\usepackage{lipsum}
\usepackage[utf8]{inputenc}
\usepackage[T1]{fontenc}
\usepackage{mathptmx}
\usepackage{amsmath}
\usepackage{color}
\begin{document}

\preprint{}

\title{The phase diagram of the Hubbard model by Variational Auxiliary Field quantum Monte Carlo}
\author{Sandro Sorella}
\affiliation{SISSA -- International School for Advanced Studies, Via Bonomea 265, 34136 Trieste, Italy;}
\affiliation{
 Computational Materials Science Research Team, RIKEN Center for Computational Science (R-CCS), Kobe, Hyogo 650-0047, Japan }
\email{sorella@sissa.it}
 

\date{\today}

\begin{abstract}
A systematically improvable  
wave function is proposed for the numerical solution of 
strongly correlated systems. With a stochastic optimization method, based on  the auxiliary field quantum Monte Carlo technique,
an effective temperature $T_{\mathrm{eff}}$ is defined,  probing the distance of  the ground state properties of the model in the thermodynamic limit from the ones of the proposed correlated mean-field ansatz.  In this way
their uncertainty  from the unbiased zero temperature limit 
may be estimated  by  
simple and stable extrapolations 
well before the so called sign problem gets prohibitive.
At finite $T_{\mathrm{eff}}$ the convergence of the energy to the thermodynamic limit is indeed shown  to be 
possible in the Hubbard model already for relatively small square lattices with linear dimension 
$L \simeq 10$,
thanks  to  appropriate averages over several twisted boundary 
conditions. 
Within the estimated energy accuracy of the proposed variational ansatz, 
two  clear phases are identified, as the energy is lowered by spontaneously 
breaking some symmetries satisfied by the Hubbard Hamiltonian: a) a stripe phase where both spin and translation symmetries are broken, and b) a strong coupling d-wave superconducting phase when the particle number is not conserved  and global $U(1)$ symmetry is broken.
On the other hand the symmetric phase is stable in a wide region at  large doping and small coupling.
\end{abstract}

\maketitle

\section{Introduction}
The accurate numerical
solution of the Schro\"edinger equation, namely determining the
ground state of a many-body Hamiltonian $H$, remains the most challenging
unsolved problem since the Dirac's formulation in 1931.
Historically, important progress, as well as too optimistic promises, occurs periodically at least twice a decade, starting from quantum computers,   the Density Functional  Theory  Kohn-Sham formulation\cite{dft}, the Density Matrix
Renormalization Group (DMRG)\cite{dmrg,dmrg_rmp} and
its translation within  the tensor network quantum  information language\cite{orus,tensor_pbc,tensor},
wrong claims\cite{me,gaussian,efetov} about  the solution of the sign
problem in Quantum Monte Carlo (QMC),
systematically improvable wave functions (WFs)  based on multi reference expansions\cite{alavi,alavi_Hubbard,umrigar_last} or on machine learning assumptions\cite{carleo,cirac}, just to mention a few of them.

Unfortunately, determining the exact
ground state of a strongly correlated Hamiltonian $H$ in the thermodynamic limit, remains an open issue 
apart for particular cases as in one dimension\cite{lieb} or with very 
particular couplings\cite{kitaev}.
Also for this reason, the Hubbard model has been historically used to benchmark new techniques, because its simplicity represents an ideal  playground 
for advanced computational and also experimental methods such as 
 optical lattices\cite{greiter}. 

As far as the 2D Hubbard model is concerned only a few results have been established by numerical techniques, namely
a clear antiferromagnetic phase at one electron per site filling\cite{hirsch,seki_benchmark,zhang_benchmark} is present and a strong evidence that this phase  remains even when this condition is not fulfilled\cite{stripe}: the added holes 
(unoccupied sites) are expelled from the antiferromagnet,  and essentially  fill in equally spaced vertical lines of the lattice: the  so called stripes.

Instead, the question of superconductivity in the Hubbard model remains highly debated and controversial since the discovery of high-temperature superconductivity\cite{cpqmc_nopair,jarrell_pair,prokovev} to the last few months\cite{absence}.
Probably the most clear evidence of superconductivity was reported in the 
strong coupling limit of the Hubbard model, the so called t-J model.
Several years ago, this model has been 
studied\cite{superb}
with an almost standard variational Monte Carlo method (VMC) that simply relies on the variational principle: the best wave function ansatz, defined by  a  set of variational parameters, is the one that minimizes the expectation value of  the  Hamiltonian studied. At that time, soon after the discovery of high-temperature superconductivity, it  was reported  that superconductivity  did  not need electron-phonon interaction but the driving  force  was rather the 
superexchange spin interaction  $J$. The  approach and especially the claim 
has been highly  debated and 
remains controversial until now, 
even though it is worth mentioning that
several years later, by  tensor network,  P. Corboz et al.  have reported,  
without emphasizing it,  almost the same value\footnote{Considering the same unit the order parameter was consistent with the previous calculation obtained by VMC\cite{superb} at around  15\% doping} of the off-diagonal $d-wave$ superconducting long range order.

Despite this success and its simplicity, the standard VMC based on a single reference 
 mean-field ansatz, e.g. the  Hartree-Fock (HF), corrected by a simple 
correlation term, e.g. the Gutzwiller factor, 
is certainly limited as compared with
the most  recent and advanced  variational methods, that  allow a systematically improvable ansatz.

In this work, a different strategy is proposed to overcome the limitations of 
correcting a simple mean-field ansatz for reaching 
accurate ground state properties. The approach 
will be dubbed in the following 
variational auxiliary field quantum Monte Carlo (VAFQMC),  that
takes advantage of the enormous 
progress done in the last decades by two old but well established techniques:
the variational quantum Monte Carlo (VMC) and the auxiliary field quantum Monte Carlo (AFQMC)\cite{thebook,umrigar,aux0,aux1,aux2}.
From the former technique its simplicity and clarity in interpreting the results in the thermodynamic limit as well as its 
ability to optimize a mean-field state $\psi_{MF}$
in presence of a simple correlated factor are taken.   
Conversely  from AFQMC  a systematically convergent and non perturbative 
expression of the correlation term is used. Indeed AFQMC allows the application of $\exp(-H\tau/2)$ to any mean-field state, where $\tau$ is the total imaginary time used in this 
formulation.
Thus, the projection $\exp(-H\tau/2)$  filters out exactly  the  ground state component of $\psi_{MF}$ for large imaginary time $\tau$, yielding that the ansatz 
$|\psi_\tau\rangle_0=\exp (-H \tau/2)| \psi_{MF} \rangle $, basically used in this work, represents  a systematically improvable- with increasing $\tau$- correlated mean-field ansatz, retaining all advantages  of the two mentioned formalisms.

The simple working  hypothesis of this study is the following. Suppose that, in  the thermodynamic limit, the mean-field wave function has acquired the lowest possible energy 
upon breaking some symmetry of the Hamiltonian in  presence of an enough accurate  correlation term. In this case,  the implicit  VMC  assumption is that the    
exact ground state should eventually show this phase. 
On the other hand, with the proposed method 
one can in principle tune the accuracy of the electron correlation 
at the desired level and  
verify explicitly 
the systematic evolution of the phase diagram, derived in this way, 
from the simplest 
mean-field HF  theory ($\tau=0$) to the converged one.
This limit can be  approached very closely 
at half-filling.
Unfortunately, for the finite doping case, the exact ($\tau=\infty$) limit cannot  be met in principle 
because the method presented here is vexed by the so called sign problem.
Nevertheless extremely accurate 
variational energies can be obtained even with 
short $\tau$ projection.  

In short the high energy degrees  of  freedom  (e.g. the average number of doubly occupied sites) of the variational ansatz  
are settled almost immediately with short $\tau$ projection (e.g. they are filtered out with the projection with an exponetial decay proportional to a large  gap $\simeq U$). 
As long as the low energy degrees of freedom 
and the high energy ones are almost decoupled, an hypothesis
implicitly assumed  in several theoretical and computational methods  in physics  and chemistry, the phase diagram derived with this ansatz should converge fast with   $\tau$, because only the high energy degrees of freedom have to be settled by the projection.
\section{Model and wave function}
In this work,  the square lattice 
Hubbard model $H= K_\mu + V$ is studied, where $K_\mu=\sum_{k,\sigma} 
c^\dag_{k,\sigma} c_{k,\sigma} (-2t (\cos k_x + \cos k_y) -\mu)$ is the 
kinetic energy,  defined here by the hopping  $t$  
and the chemical potential $\mu$ and $V= U \sum_i n_{\uparrow,i} n_{\downarrow,i}$ is the total number of doubly occupied sites operator 
scaled by the coupling $U>0$ of the model,
where standard  second quantization notations are assumed.
We introduce the following ansatz\cite{Baer1,nippo1,nippo2}, 
also inspired by similar wave functions quite popular in quantum computation\cite{melko,troyer}, 
with the purpose to reach accurate ground state properties in the shortest possible projection 
time $\tau$:
\begin{equation} \label{wf}
|\psi_\tau \rangle = \exp \left[ -{\tau \over  2} (H_{MF}(\bm{\alpha}) + V)\right] |\psi_{MF} \rangle.
\end{equation}

As shown later, the above variational wave function represents a straightforward improvement of the standard auxiliary  field QMC one $|\psi_\tau \rangle_0$,  mentioned in the introduction. 
The mean-field wave function $|\psi_{MF} \rangle$ can be  a generic 
quasi-free electron state, from a simple Slater determinant to BCS pairing 
functions, including singlet and/or triplet correlations.
Here, $|\psi_{MF} \rangle$
 is allowed breaking some symmetries of the Hamiltonian, and is defined 
as the ground state of a mean-field Hamiltonian $H_{MF}(\bm{\alpha_0})$, i.  
e. $H_{MF}(\bm{\alpha_0}) |\psi_{MF} \rangle  = E_0(\bm{\alpha_0}) |\psi_{MF} \rangle$ with a set  of  variational parameters indicated by the vector 
$\bm{\alpha_0}$.
For the projection operator,  we take the  advantage of the variational formulation, so that any 
extension of the variational ansatz containing a given one- i.e. $|\psi_\tau \rangle_0$-  as a particular 
case should necessarily improve it, namely after energy optimization,  
acquires a lower variational  energy.
The bare kinetic energy operator in $\exp(-H \tau/2)$ is therefore generalized (i.e. $K_\mu \to H_{MF}(\bm{\alpha})$), 
by allowing  
a generic operator quadratic in the fermion ones $c,c^\dag$, including 
for instance also a  d-wave BCS pairing field.  
$H_{MF}(\bm{\alpha})$ is therefore parameterized by
a set 
of variational parameters indicated 
by another independent vector $\bm{\alpha}$, such that for $\bm{\alpha}=0$ 
$H_{MF}(\bm{\alpha})=K_\mu$. In other words the idea is that if a symmetry is 
broken in the thermodynamic limit also the projection operator, i.e. the correlation factor,  and not only the 
mean-field state $|\psi_{MF}\rangle$ may break the symmetry and therefore $H_{MF}(\bm{\alpha})$  is conveniently parametrized in a way similar to $\psi_{MF}$.

This original formulation 
 improves the quality of the ansatz 
 and the smooth convergence to the thermodynamic limit, as compared with the simpler ansatz $|\psi_\tau \rangle_0$ and
the optimal energy is therefore generally obtained after  the simultaneous optimization of both   $\bm{\alpha}$ and $\bm{\alpha_0}.$

We anticipate that, in this variational formulation of the auxiliary field 
quantum Monte Carlo  (VAFQMC), $\tau$ plays the 
role of an effective inverse 
temperature that is kept fixed during the minimization 
of  the energy expectation value corresponding to $|\psi_\tau \rangle$.  

 The optimization techniques known in standard variational Monte Carlo\cite{1998SOR} and machine learning\cite{ml1} will be generalized here to
the auxiliary field  QMC. 
Before that,  it is worth  to emphasize  simple but important properties 
of this ansatz:
\begin{enumerate}
\item It is systematically improvable. In order to realize this property, it is enough to take  
$\bm{\alpha}=0$, when the 
proposed ansatz coincides with the simpler one $|\psi_\tau\rangle_0=
\exp(-H \tau/2)|\psi_{MF}\rangle$. 
Thus,  let $\tau\to \infty$ for a  
mean field $|\psi_{MF}\rangle$ of the chosen form, that  is 
not orthogonal to the ground state. 
In  this limit 
$|\psi_\tau\rangle_0$ 
is obviously converging to the exact ground state.  
Thus, after turning on optimization, 
by allowing both $\bm{\alpha} \ne 0$ and, independently, $\bm{\alpha_0}$ different 
from the initial guess, 
a lower energy is necessarily implied 
for each $\tau$, yielding that the ansatz of Eq.~\ref{wf} is 
systematically improvable, as well as and even better than  $|\psi_\tau\rangle_0$ as far as its energy is concerned.
\item It is size extensive. As discussed in Ref.\onlinecite{thebook} 
the WF is defined directly in terms of an exponential of an extensive 
operator, hence the statement. 
In  practice this means 
that, at a given $\tau$, approximately  
the same accuracy for  intensive quantities is 
expected, e.g.  the energy per site or bulk  correlation functions.
\item For finite clusters the convergence is exponential in $\tau$ due to the finite size gap between the ground state manifold (which may be  also degenerate)
and the first excitation with non zero energy gap. However in the thermodynamic limit this gap is probably  always vanishing in this model, even for 
the half filled insulator,  where gapless spin-wave excitations are expected due to the occurrence of 
antiferromagnetic order for any  $U>0$\cite{seki_benchmark}.
This situation,  as it  will be shown in the  following, makes the extrapolation to 
the ${1\over \tau} =0$ unbiased limit much simpler than the corresponding 
finite size case. Indeed this can be obtained
by simple and stable low order 
polynomial extrapolations in something that can be considered 
an effective temperature $T_{\mathrm{eff}}=1/\tau$. 
\end{enumerate}
The key idea of this work is to converge   first the results to
the thermodynamic limit with large enough 
cluster size simulations and appropriate  boundary  conditions. 
Given this,  simple and very stable 
extrapolations in $T_{\mathrm{eff}}\to 0$ are employed,  thus achieving, 
with  this simple minded strategy, very accurate  results of the model or at least an estimate of the accuracy of the lowest $T_{\mathrm{eff}}$ variational ansatz.
As presented later, this is often possible because,  after the  optimization,  the  physical properties of the ansatz given  in Eq.(\ref{wf}) are 
already in  the very low temperature regime, where the mentioned extrapolations are  indeed stable and reliable within a given broken symmetry phase.


It is well established\cite{fisher,hasenfratz,hofmann}  that, when a  standard type of 
order that breaks a continuous symmetry sets in, the corresponding 
gapless low energy 
excitations (i.e. typically  bosons with a density 
of states $\rho(\epsilon)\propto \epsilon^{D-1}$)  induce $\simeq T^3$ ($\simeq T^2$)  energy  corrections (i.e. $\Delta  E \propto \int\limits_0^T \epsilon \rho(\epsilon)\simeq T^{D+1}$) in the limit $T\to 0$ for 2D (quasi 1D system like a finite 
cylinder with infinite length), and indeed it turns out that the  fit 
\begin{equation} \label{chosenfit}
E(T_{\mathrm{eff}}) =E(0)+ a T_{\mathrm{eff}}^{D+1} + b T_{\mathrm{eff}}^{D+2}
\end{equation}
is generally very appropriate 
for the systems studied here, because the available  $T_{\mathrm{eff}}$ appear always 
quite close to 
the correct asymptotic behavior. 
A formal derivation of the finite effective temperature corrections is given in App.\ref{sec:ftemp}.

Given the above arguments,   
particularly useful boundary conditions are adopted  
such that the convergence to the 
thermodynamic limit at fixed $\tau$ is as fast as possible.
To this purpose twisted averaged boundary conditions TABC\cite{natoli,
gros} are used in both the Hubbard Hamiltonian 
and the mean-field ones on rectangular 
$L_x \times L_y$ clusters ($L_x=L_y=L$ is adopted for square lattices).
 This is obtained  by imposing opposite twists
 for opposite spins:
 \begin{eqnarray}
 c_{r_x+L_x,r_y,\uparrow}&=&c_{r_x,r_y,\uparrow}\exp( i 2\pi  \theta_x ) \nonumber \\
 c_{r_x,r_y+L_y,\uparrow}&=&c_{r_x,r_y,\uparrow}\exp( i 2\pi  \theta_y ) \nonumber \\
 c_{r_x+L_x,r_y,\downarrow}&=&c_{r_x,r_y,\downarrow}\exp( -i 2\pi  \theta_x ) \nonumber \\
 c_{r_x,r_y+L_y,\downarrow}&=&c_{r_x,r_y,\downarrow}\exp(-i 2\pi  \theta_y ) \nonumber \\
 \end{eqnarray}
 with $\theta_x=-1/2 + (i-1/2)/N_T$ and $\theta_y=-1/2+(j-1/2)/N_T$
 for integers $1\le i,j \le N_T$, while $r_x$ and $r_y$ here and henceforth indicate integer Cartesian coordinates of the lattice $1\le r_x\le L_x$, $1\le r_y \le L_y$.
All the results are then  
averaged 
on a mesh of $N_T\times N_T$  twists in the Brillouin zone (BZ), with  $N_T$ large enough 
to have  converged energies within statistical  errors. 
When $H_{MF}$ conserves the number of particles $N$ and has a gap in the one particle spectrum, $N$ remains unchanged for each twist, whereas  when a BCS pairing  is present,
grand canonical ensemble is adopted as discussed in  Ref.\onlinecite{karakuzu} 
and the expectation value of $H-\mu \hat  N$ is minimized by the proposed variational ansatz, where $\hat N=\sum\limits_{r_x,r_y,\sigma} c^\dag_{r_x,r_y,\sigma} c_{r_x,r_y,\sigma}$ is the particle number operator.

The use of opposite twists for opposite spin electrons is particularly 
important in this case because it allows the conservation of the  translation 
symmetry in the  BCS mean-field Hamiltonian,  at least in a simple way.

\begin{figure}[htbp]
  \centering
  \includegraphics[width=8.6cm]{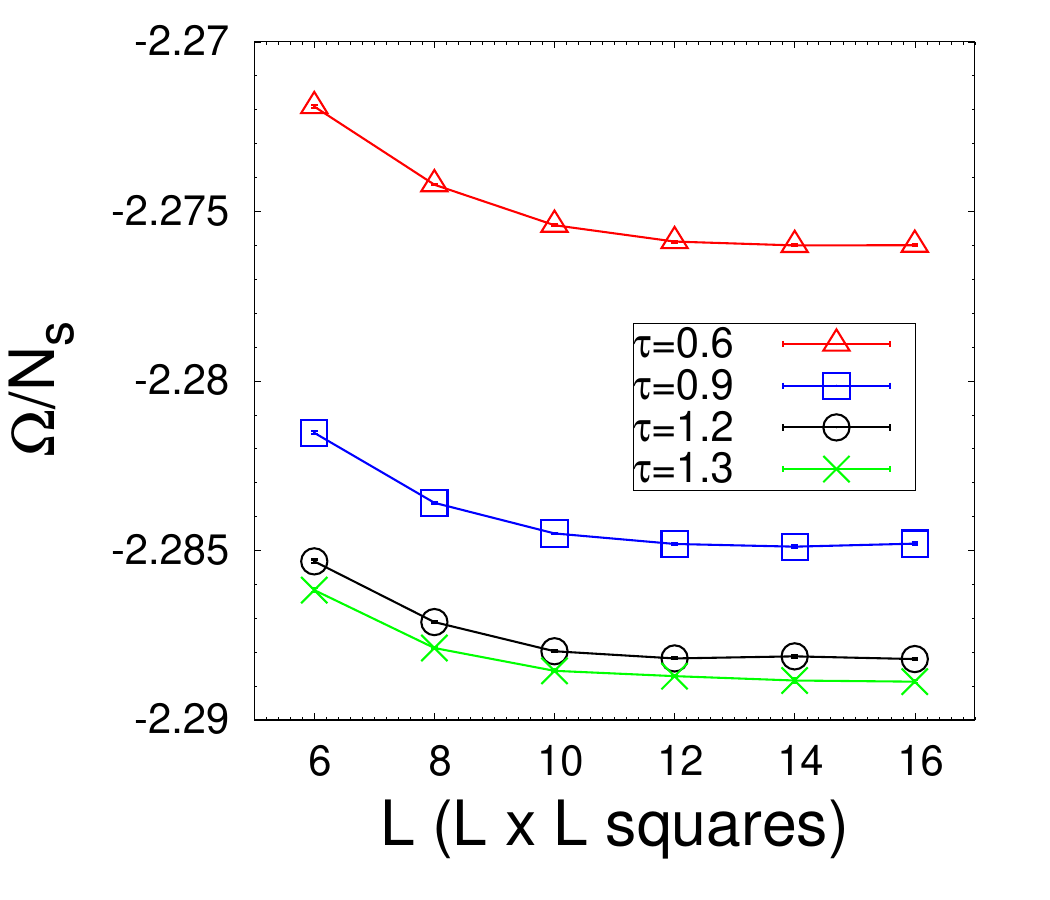}
    \caption{Grand potential  per site $\Omega/N_s$ as  a function of cluster size for  different 
    values  of the projection  time  $\tau$  for $\mu=1.75t$ and $U/t=8$}
\label{ebvslfig}
\end{figure}

Before explaining how to optimize the variational parameters it is useful 
to appreciate in Fig.~\ref{ebvslfig} 
the fast and smooth  convergence of the grand potential  
$\Omega={\langle \psi_\tau| H-\mu \hat N| \psi_\tau\rangle \over 
 \langle \psi_\tau| \psi_\tau\rangle }$  in  the thermodynamic limit as a function of the  number $N_s$ of  
sites 
 for a value of the chemical potential corresponding to doping  
$\delta= 1-N/N_s \approx 1/8$.

The other important ingredient for an efficient implementation of 
Eq.~(\ref{wf}) is 
the use of a  particularly suited Trotter decomposition for  the 
corresponding propagator:
\begin{eqnarray} \label{trotter}
&\exp \left[ -{\tau \over  2} (H_{MF}(\bm{\alpha}) + V)\right]=& \nonumber \\
& \left\{ \prod\limits_{i=1}^n \exp \left[ - t_i H_{MF}(\bm{\alpha}) \right] 
\exp\left[ - h_i V \right] \right\} \times &  \nonumber \\
&  \exp\left[ - t_{n+1} H_{MF}(\bm{\alpha}) \right] & 
\end{eqnarray}
where in principle $h_i$ and $t_i$ can be  independently optimized 
to minimize the Trotter error,
as proposed in  a recent work\cite{melko}.
Henceforth it is assumed that  the operators in Eq.~(\ref{trotter})
are ordered from left to right according to increasing values of 
the integer $i$.
In this work $\tau$ is kept fixed,
 and therefore the following  constraint is imposed:
\begin{equation} \label{eqconstr}
\sum\limits_{i=1}^n h_i= {\tau \over 2}.
\end{equation}
Moreover, in order to minimize the number of variational parameters  
and the number $n$,  
the following  parameterization is adopted:
\begin{eqnarray}
h_i &=&  \Delta \tau \gamma^{i-1} \label{defdt} \\ 
t_i &=& { h_i + h_{i-1} \over 2}  ~~{\rm for} ~ i=1,2,\cdots n
\end{eqnarray}
with $h_{0}=0$ and  $\gamma>1$ chosen in a way to satisfy the constraint in Eq.~(\ref{eqconstr}).
This expression is based on the conventional small $\Delta \tau$ 
symmetric Trotter decomposition $\exp \left[ -{\Delta \tau } (H_{MF}(\bm{\alpha}) + V)\right] = \exp \left[ -{\Delta \tau \over  2} H_{MF}(\bm{\alpha})\right] 
\exp \left[ -\Delta \tau V\right] \exp \left[ -{\Delta \tau \over  2} H_{MF}(\bm{\alpha}) \right] +O(\Delta\tau^3)$, where a nonuniform 
time step according to  Eq.~(\ref{defdt}) has been  chosen 
because, for high  accuracy, it is important only that 
the first time step is small and $\gamma \simeq 1$.
In this way the  
 efficiency and the number of operators involved is significantly optimized, 
 without using too many variational parameters as in  Ref.\onlinecite{melko}.
In all forthcoming calculations  only a single parameter $\Delta \tau$ is optimized 
with the convenient choice:
\begin{equation}
n={\rm Max} \left\{  [(\tau U/0.4-1)/5],1 \right\},
\end{equation}
leading to a Trotter time step error on the energy  that, for all $U$ values reported, can be hardly distinguished from the error bars. Here and henceforth the square brackets indicate the integer part of a real number.

\section{Variational Auxiliary Field Method}
Since $H_{MF}(\bm{\alpha})$ is a one body operator and $|\psi_{MF}\rangle$ is a
mean-field state 
(e.g.  a Slater determinant) there is no difficulty to apply $\exp\left[-t_i H_{MF}\bm(\alpha)\right]$  to $|\psi_{MF}\rangle$. Moreover,  taking into account that
$\exp\left[-t_i H_{MF}\bm(\alpha)\right| \psi_{MF}\rangle$ 
remains a mean-field state, this operation can be also performed iteratively.

The application of $\exp( - h_i V) $ is instead more complicated and can nevertheless be implemented by using the auxiliary field technique. 
To this purpose the following discrete Hubbard-Stratonovich transformation\cite{hirsch} is used:
\begin{equation}
\exp( -  h_i V) = 2^{-N_s} \exp(-U h_i \hat N/2)  \sum\limits_{\sigma_{j,i}=\pm 1} \exp( \lambda_i \sum\limits_j \sigma_{ji} m_j)
\end{equation}
where $m_j= n_{j,\uparrow}-n_{j,\downarrow}$, $\hat N$  is the total number of particles operator, $\cosh \lambda_i= \exp( U h_i)$ and the Ising variables $\sigma_{ji} = \pm 1$  for each site $j$ of the lattice. Notice that  the factor $\exp(-U h_i \hat N/2)$  
represents only an irrelevant  change  of  the chemical potential in $H_{MF}(\bm{\alpha})$. Therefore this term is omitted for simplicity here and henceforth.

We  get  therefore that the expectation value of  Hamiltonian $H$ (or any relevant correlation function\footnote{From this point of view VAFQMC is not different from standard AFQMC or VMC and the  expectation value of any operator, not only $H$,  can be computed without particular effort.\cite{thebook}}) for the wave  function in Eq.(\ref{wf}) is given by:
\begin{eqnarray}
E_n&=& {\langle \psi_n| H| \psi_n\rangle \over  \langle \psi_n| \psi_n\rangle } 
\label{eqpsi} \\
&=& { \sum\limits_{\bm{\sigma} \bm{\sigma^\prime}}  \langle \psi_{MF} | U^\dag_n(\bm{\sigma^\prime}) H U_n(\bm{\sigma}) | \psi_{MF}  \rangle \over 
 \sum\limits_{\bm{\sigma} \bm{\sigma^\prime}} \langle \psi_{MF} | U^\dag_n(\bm{\sigma^\prime})  U_n(\bm{\sigma}) | \psi_{MF}  \rangle }
\end{eqnarray}
where $\bm{\sigma}$ indicates the $N_s\times n$ dimensional vector with  
components $\sigma_{ji}$, and: 
\begin{eqnarray}
& U_n(\bm{\sigma})& =  \exp(-H_{MF}(\bm{\alpha}) t_1) \exp( \lambda_1 \sum_j \sigma_{j,1} m_j )   \\
&\cdots & \exp(-H_{MF}(\bm{\alpha}) t_n) \exp( \lambda_n \sum_j \sigma_{j,n} m_j) \exp(-H_{MF}(\bm{\alpha}) t_{n+1}) \nonumber 
\end{eqnarray}
$E_n$ can be therefore computed by Monte  Carlo, by sampling the Ising fields 
$\bm{\sigma}$ and $\bm{\sigma^\prime}$ according to the  weight $|W_n(\bm{\sigma^\prime},\bm{\sigma})|$, where:
\begin{equation}
W_n(\bm{\sigma^\prime},\bm{\sigma}) = \langle \psi_{MF} | U^\dag_n(\bm{\sigma^\prime})  U_n(\bm{\sigma}) | \psi_{MF}  \rangle 
\end{equation}
and, quite generally in the complex case, 
$W_n$ has a non trivial phase that  is determined by   $S_n(\bm{\sigma^\prime},\bm{\sigma}) =  { W_n(\bm{\sigma^\prime},\bm{\sigma}) \over | W_n(\bm{\sigma^\prime},\bm{\sigma})| }$, 
a complex number with unit modulus, that plays the  role of the QMC 
sign.

Finally $E_n$ can be computed by :
\begin{equation} \label{enp}
E_n= {\sum\limits_{\bm{\sigma} \bm{\sigma^\prime}} |W_n(\bm{\sigma^\prime},\bm{\sigma})| e_n(\bm{\sigma^\prime},\bm{\sigma}) S_n(\bm{\sigma^\prime},\bm{\sigma}) \over 
 \sum\limits_{\bm{\sigma} \bm{\sigma^\prime}} |W_n(\bm{\sigma^\prime},\bm{\sigma})|  S_n(\bm{\sigma^\prime},\bm{\sigma})  
}
\end{equation}
namely by evaluating the ratio of the means corresponding to two real random variables $ \Re[e_n(\bm{\sigma^\prime},\bm{\sigma}) S_n(\bm{\sigma^\prime},\bm{\sigma})]$ and $\Re[S_n(\bm{\sigma^\prime},\bm{\sigma})]$ 
\footnote{the numerator and the denominator of Eq.~(\ref{enp}) are both real after summations over $\sigma$ and $\sigma^\prime$ because they correspond to the RHS of Eq.~(\ref{eqpsi}) and the Hamiltonian is Hermitian.
Therefore the real part $\Re$  can be moved  inside the summations of Eq.~(\ref{enp}) with  no approximation.}
over the configurations generated by Monte Carlo sampling according to  the  probability 
$p_n(\bm{\sigma^\prime},\bm{\sigma})={|W_n(\bm{\sigma^\prime},\bm{\sigma})| \over \sum\limits_{\bm{\sigma} \bm{\sigma^\prime}} |W_n(\bm{\sigma^\prime},\bm{\sigma})|}$, using 
the standard technique described in Ref.\onlinecite{thebook}, and:
\begin{equation}
e_n(\bm{\sigma^\prime},\bm{\sigma})= {
\langle \psi_{MF} | U^\dag_n(\bm{\sigma^\prime}) H  U_n(\bm{\sigma}) | \psi_{MF}  \rangle  \over W_n(\bm{\sigma^\prime},\bm{\sigma})  } 
\end{equation}
is a  sort of local energy, namely an estimate of $E_n$ for a given  configuration  of the  Ising  fields $\bm{\sigma}$ and $\bm{\sigma^\prime}$.
Indeed both $e_n(\bm{\sigma^\prime},\bm{\sigma})$ and $S_n(\bm{\sigma^\prime},\bm{\sigma})$, as well as $W_n(\bm{\sigma^\prime},\bm{\sigma})$ can be 
computed in $\propto  N_s^3 n$ operations because they involve essentially 
imaginary time propagations of mean-field states under time dependent 
one body propagators $U_n(\bm{\sigma})$ and $U_n(\bm{\sigma^\prime})$.

The basic ingredient introduced in this work is the possibility to
compute energy derivatives of $E_n$ with respect to all the parameters defining the WF, i.e. $\bm{\alpha}$ and $\bm{\alpha^\prime}$, henceforth assumed to be defined by the 2p variational parameters $\alpha_1,\alpha_2,\cdots \alpha_{2p}$ 
and the minimum time  step used $\Delta \tau=\alpha_{2p+1}$ at fixed $\tau$, according to Eq.(\ref{eqconstr}).  A simple algebra, very similar to the one 
known for VMC, implies that any energy derivative  ${\partial
E_n\over \partial \alpha_j} $ 
with  respect to an  arbitrary variational  parameter $\alpha_j$, for $j=1,2,\cdots 2p+1$, can be computed by means of
corresponding derivatives of two complex functions  ${\partial e_n \over \partial \alpha_j}$, 
$O_j= { \partial \ln( W_n) \over \partial \alpha_j} =
{\partial W_n \over \partial \alpha_j}/W_n$  
of the  local energy and 
the logarithm of the weight, respectively:
\begin{equation} \label{defsign}
{\partial E_n \over \partial \alpha_j} = { \langle \langle  \Re\{ S_n \left[{\partial e_n \over \partial \alpha_j}  +(e_n-E_n)  O_j \right] \} \rangle \rangle  \over 
\langle \langle   \Re(S_n) \rangle \rangle }
\end{equation}
where here and henceforth the symbol $\langle \langle *\rangle \rangle$ indicates the average 
of the generic random variable $*$ over the probability  distribution 
$p_n$ defined before, and for shorthand  notations the dependence 
on $\bm{\sigma}$ and $\bm{\sigma^\prime}$ of all the quantities involved 
is not explicitly  shown.

The differentiation of the complex quantities $\ln W_n$ 
and $e_n$, required for the ${\partial E_n\over \partial  \alpha_j}$ evaluation, at given 
values of $\bm{\sigma}$  and $\bm{\sigma^\prime}$,
may appear very 
cumbersome and involved especially  considering that, it  is often necessary, 
as in VMC, to optimize several parameters. This task can be easily  achieved
in a computational time equal to  the one required  to compute 
the complex  quantities $\ln(W_n)$ and $e_n$, remarkably only 
up to a small prefactor 
 regardless how large  is the  number of variational parameters involved. 
This is possible by using Adjoint Algorithmic 
Differentiation (AAD)\cite{capriotti_fin_old}, a technique that is becoming popular 
in the field of Machine Learning with another name, i.e. 
''back propagation'', but was certainly known before in 
applied mathematics\cite{capriotti_fin},  and  only  
recently has been appreciated in physics\cite{capriotti,wang}.
Once all energy  derivatives $\partial_j E $ 
are known, the usual scheme adopted in VMC can be applied also 
here. Variational parameters are changed according to the equation:
\begin{equation}
 \delta \bm{ \alpha} = - {\rm rate_{learning}}  F^{-1}  {\partial  E \over 
 \partial \bm{\alpha}} 
\end{equation}
 where ${\rm rate_{learning}}$ is a suitable small constant, determining the 
 speed of convergence to the minimum and $F$ 
 is the so called Fisher-information matrix, given by:
 \begin{equation}
 F_{ij} = \langle \langle { \partial \log p_n \over \partial \alpha_i} 
 {\partial  \log p_n \over \partial \alpha_j} \rangle \rangle=\langle \langle\langle \Re(O_i) \Re(O_j)\rangle \rangle \rangle
 \end{equation}
where the symbol $\langle \langle \langle  A B \rangle \rangle \rangle = \langle \langle  A B  \rangle \rangle  - \langle \langle  A\rangle \rangle  \langle \langle  B\rangle \rangle $ indicates here the covariance of two random variables 
over  the probability $p_n$. We adopt here ${\rm rate_{learning}} \simeq 6$ 
and the same regularization with  $\epsilon=0.01$ described 
in  Ref.~\cite{thebook} to avoid instabilities in  the calculation of the inverse matrix. Typically convergence is reached with  a few hundreds iterations
and, variational  parameters are averaged after convergence for about $50$ steps.
\section{Testing the method}
In this section the proposed  method is tested 
against known benchmark results on 
infinite systems.
Unfortunately there are only a few results available in the thermodynamic limit, and those ones are mainly limited to ground state energies. Nevertheless they are extremely relevant for a variational method, because its predictions can be supported by a good estimate of the energy.

\begin{figure}[htbp]
  \centering
  \includegraphics[width=8.6cm]{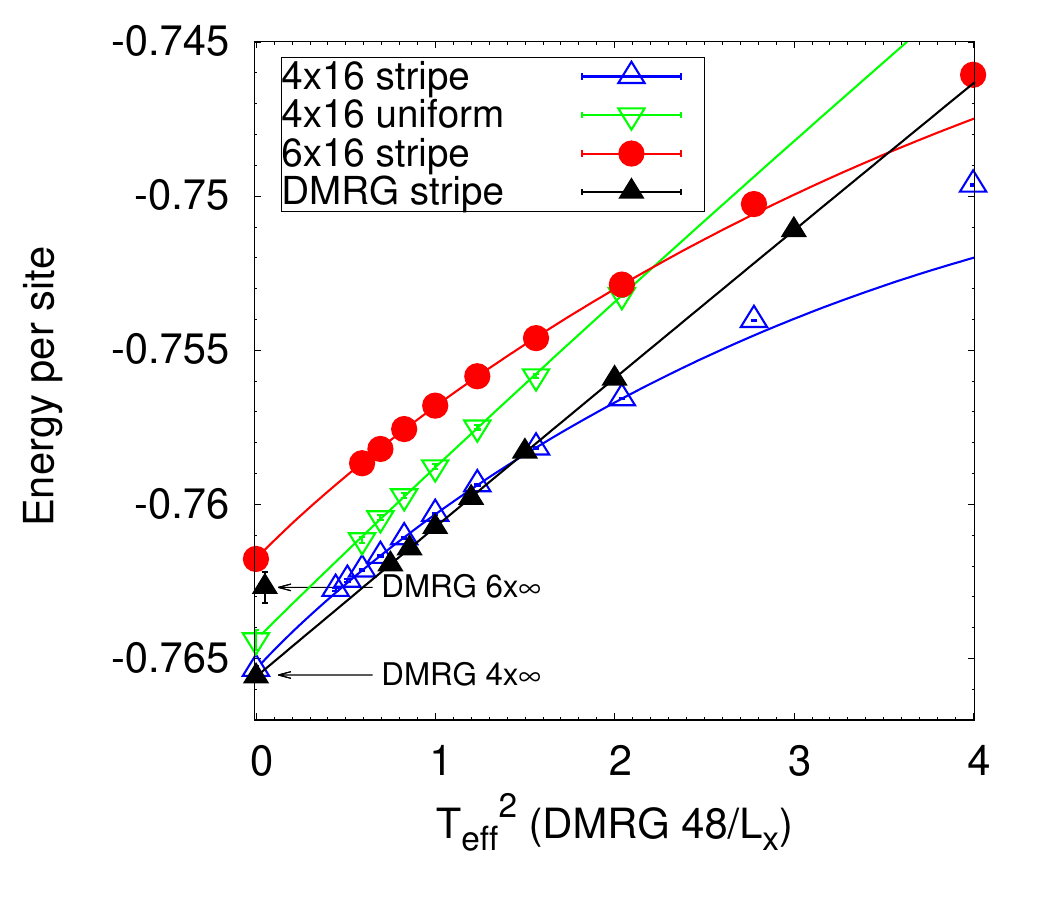}
    \caption{  Energy as a function of  the effective temperature  $T_{\mathrm{eff}}=1/\tau$  for cylinders with different widths. Values at zero horizontal axis result from  extrapolations.
    Here the $D=1$ asymptotic form of Eq.~\ref{chosenfit} is adopted for the fit of the data in the range  $0.7\le \tau t \le 1.3$. This form is compatible with the expected low energy  spectrum (see text) while the range used is found to be appropriate for $U=8t$  in all cases, included the 2D ones, studied in this work. DMRG energies obtained with open boundary conditions are reported  as a function of the inverse  cylinder length $L_x$  for the $4\times L_x$ case, while  the extrapolated ones at $L_x=\infty$ for both  the $4 \times L_x$ and $6\times L_x$ cases.}
\label{ebvstfig}
\end{figure}

First VAFQMC is tested at $U/t=8$ 
on cylinders with finite width $L_y$ and  periodic 
boundary conditions (PBC) in 
the short $y-$ direction, in order to compare
with the very accurate 
results determined by DMRG with open boundary conditions (OBC) in the $x-$direction, 
and extrapolated to the one dimensional  $L_x=\infty$
thermodynamic limit. As it is seen in Fig.~\ref{ebvstfig} VAFQMC is in very good agreement with  the known results for $L_y\le 6$, especially when the $T_{\mathrm{eff}}^2 \to 0$ 
extrapolation is employed in a relatively small $L_y\times 16$ cluster with $32$ 
twists in  the long direction\footnote{convergence in the thermodynamic limit has also been verified  by comparing with the $4\times 32$ cluster within an error of 
about $0.0001t$}.
In this picture we can also appreciate the importance to use appropriate boundary  conditions to reach the thermodynamic limit, as OBC require very  large clusters for this purpose, a clear limitation  that prevents a very accurate 
extrapolation. Notice that VAFQMC extrapolated results, though affected by a visible non linear term, are obtained with variational energies that are better 
(below) than the available\footnote{Obviosuly possible DMRG calculations with $L_x\simeq 100$ should provide energies better than the VAFQMC reported ones. So this remark does not imply that VAFQMC provides in general variational energies better than DMRG.}  DMRG ones also for the smallest width $L_y=4$. This is remarkable because  DMRG has obviously best performances in  almost  one dimensional systems. 
The small discrepancy  between VAFQMC and DMRG extrapolated values for $L_y=6$ does not necessarily imply that VAFQMC is less accurate in  this case, but may be  due to  the  limitation of DMRG in obtaining reliable extrapolations as soon as we approach the 2D case, as also shown by the large error bar reported. In any case it is clear that the proposed extrapolations, can be used to estimate the quality of the VAFQMC best (lowest $T_{\mathrm{eff}}$) 
variational energy estimates.   

\begin{figure}[htbp]
  \centering
      \includegraphics[width=9cm]{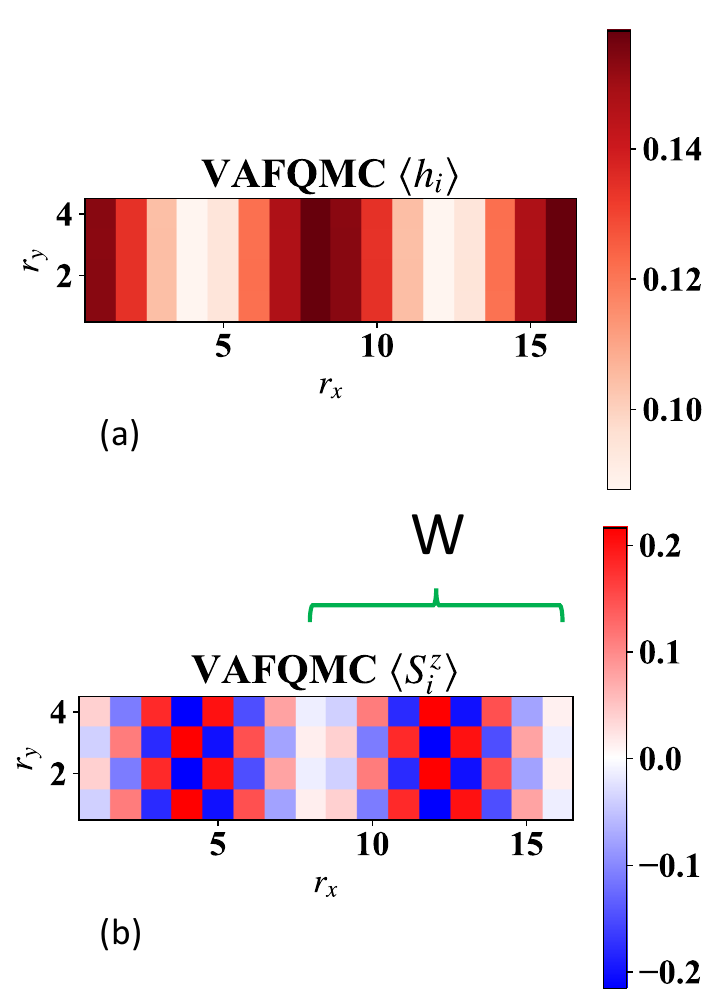}
           \caption{Hole density (a) and spin density (b) for the stripe of width $W=8$ and $U/t=8$ for the $4\times 16$ cluster  obtained by energy optimization at $\tau t=1.5$. The green bracket emphasizes the length $W$ of the stripe, i.e. the distance between two domain walls.} 
        \label{stripe_4x16}
 \end{figure}

In  all these VAFQMC calculations, at doping $\delta=1/8$ 
it  has been found that the $W=8$ 
stripe is the most favorable mean-field, as shown in Fig.~\ref{stripe_4x16} for the $L_y=4$ case.
Here this state is parametrized by the 
most general 
mean-field Hamiltonian with local and nearest neighbor couplings 
independent of $r_y$ (see also the Supplementary Information): 
\begin{eqnarray}
H_{MF}(\bm{\alpha_0}) &=& \sum\limits_{r_x,r_y,\sigma} \left[-t_x(r_x) c^\dag_{r_x+1,r_y,\sigma} c_{r_x,r_y,\sigma}\right. \nonumber \\
&-& \left. t_y(r_x) c^\dag_{r_x,r_y+1,\sigma} c_{r_x,r_y,\sigma} + {\rm h.c.} \right] \nonumber \\
&+& \sum\limits_{r_x,r_y} \Delta_{AF}(r_x) (-1)^{r_x+r_y} (n_{\uparrow,r_x,r_y}-n_{\downarrow,r_x,r_y}) \nonumber \\
&-&\mu(r_x) n_{r_x,r_y}
\end{eqnarray}
where $n_{\uparrow,r_x,r_y}=c^\dag_{\uparrow,r_x,r_y} c_{\uparrow,r_x,r_y}$,
$n_{\downarrow,r_x,r_y}=c^\dag_{\downarrow,r_x,r_y} c_{\downarrow,r_x,r_y}$,
$n_{r_x,r_y}=n_{\uparrow,r_x,r_y}+n_{\downarrow,r_x,r_y}.$
Thus 
the  local magnetic antiferromagnetic field
satisfies $\Delta_{AF}(r_x)=-\Delta_{AF}(r_x+W)$
and the corresponding local chemical potential $\mu({r_x})=\mu(r_x+W)$.
Moreover 
 nearest neighbor hoppings $t_x(r_x)=t_x(r_x+W)$ and 
$t_y(r_x)=t_y(r_x+W)$ are also included, amounting to a total of $4W$ 
variational parameters 
defining $H_{MF}(\bm{\alpha_0})$.
All their values 
are assumed independent of the twists\cite{karakuzu}.
The few parameter choice 
adopted in Ref.\onlinecite{tocchio_stripes} is used only to initialize 
the present  most general  $8W+1$ 
(by  including  also the optimization of $H_{MF}(\bm{\alpha})$ and $\Delta \tau$) parameter ansatz $|\psi_{\tau}\rangle$.

For the uniform solution (see Sec.~\ref{sec:corrps}) antiferromagnetic order is allowed only  at  half filling while at finite doping  a four (three)  
parameter ansatz is adopted 
in $H_{MF}(\bm{\alpha})$ ($H_{MF}(\bm{\alpha_0})$, here $t=1$ can be left unchanged, as it
sets the scale of the mean-field energy, irrelevant for $|\psi_{MF}\rangle$) 
including nearest and next nearest 
neighbor hoppings as well has a uniform chemical potential $\mu_0$ and 
$d_{x^2-y^2}$ pairing.
In all the forthcoming calculations, when the particle number is not defined neither in the mean-field wave function nor in the projection,  the energy per site $e(\delta)$  at fixed doping $\delta$ is accurately estimated by an appropriate choice of the chemical  potential $\mu$.  This is  obtained by simple and stable interpolations with a few calculations in the grand canonical ensemble or, in other words,  by  inverting the Legendre transform from chemical  potential dependence to the conjugate density one $1 -\delta$. 

With this method it is therefore possible to compute the energy in  the thermodynamic limit of true 2D clusters   without particular effort as the average (complex-) sign 
$\langle \langle  S_n\rangle \rangle $ is always larger than $\simeq 0.3$ for all the simulations reported in this work. It is important to remark that, when  
the 2D-thermodynamic limit is approached, a small but systematic increase of the energy 
can  be appreciated, confirming that, in order to determine the correct two-dimensional thermodynamic limit, true two-dimensional clusters have to be used.

For $L_y$ small the stripe solution is clearly favored as compared with the uniform solution, as found by DMRG. In Fig.~\ref{ebvstfig} the uniform 
solution is optimized with a non zero $d_{x^2-y^2}$ BCS pairing but remains  clearly above the stripe solution even after extrapolation to $T_{\mathrm{eff}}=0$.

However the situation is quite different when the 
two-dimensional thermodynamic limit is required. 
In this case, in order to compare the 
stripe with periodicity $W=8$ and $W=7$
and the uniform $d-wave$ solution, calculations on a
$16\times 16$ and  $14\times 16$ clusters
are carried out with $16\times16$ different twists in the BZ. For  the $d-$wave ansatz several calculations at different chemical potentials  $\mu \simeq 1.8t$ are attempted in order  to fulfill a doping $\delta = 1/8$ for each effective temperature $T_{\mathrm{eff}}$.

\begin{figure}[htbp]
  \centering
      \includegraphics[width=8.6cm]{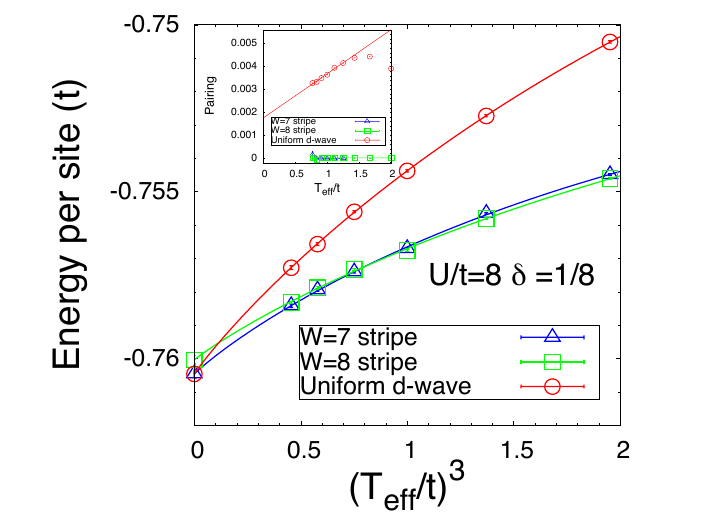}
              \caption{ Energy per site
	      at doping  $\delta =1/8$  and $U/t=8$ for the stripe solution of different lengths and
           the uniform $d-wave$ superconducting one, corresponding to the same doping. Values at zero horizontal axis result from  extrapolations.
	   Here the $D=2$ fitting form of Eq.~\ref{chosenfit} is adopted, that is compatible with the expected low energy  spectrum (see text).  A $16\times 16$ square lattice was used for the $W=8$ stripe and the uniform $d-$wave states while a $14\times 16$ rectangular lattice for the $W=7$ stripe. The inset shows the pairing correlations at maximum horizontal distance $L_x/2.$}
        \label{16rung}
 \end{figure}

It is seen from the Fig.~\ref{16rung} that in 2D the uniform solution turns out to have an energy  competitive with the stripe ones, though the $W=7$ stripe appears to be the most 
stable solution also because it provides always the lowest variational energy for 
each $T_{\mathrm{eff}}$. 
Nevertheless  the stripe-uniform phase  energy difference, reached at the lowest 
$T_{\mathrm{eff}}$ shown, is very small  $\simeq 0.001t$, though the two 
variational states remain with qualitatively different types of correlation functions, i.e. a sizable magnetic moment in the former and non negligible $d-wave$ pairing correlations in the latter (see inset in Fig.~\ref{16rung}).
In Tab.~\ref{tab_compare} the various extrapolated and best 
variational energies  are compared with the  available  benchmark results\cite{zhang_benchmark,stripe,tocchio_stripes} obtained with other variational methods.
Remarkably VAFQMC provides always the best known variational energies in the 
thermodynamic limit. 
In 2D the VAFQMC performances are manifestly excellent if compared with other variational methods, because DMRG becomes rapidly inaccurate with increasing $L_y$, and 
the best variational energy obtained by iPEPS is significantly higher than the VAFQMC one. 
This  may explain why the iPEPS 
extrapolated energy has a large uncertainty and is much below the VAFQMC and DMRG $6\times \infty$ ones.
It is also possible that the extrapolation in  the bond dimension  is 
quite inaccurate, and can be substantially improved\cite{TagliacozzoX}. 
\begin{table}[h]
  \begin{tabular}{|l|l|l|l|}
      \hline
     {\rm Method}  &  $4\times \infty$  &  $6\times \infty$ & $\infty \times \infty$    \\
      \hline
{\rm VMC+backflow }\cite{tocchio_stripes}  & ==  &  -0.7483(1)  & -0.74884(1) \\
{\rm   DMRG}$^*$\cite{stripe}   &  -0.76598(3) & -0.7627(5) & ==   \\
{\rm   DMRG}$^*$\cite{private_dmrg}  & -0.7655(1)  & == & ==   \\
{\rm   DMRG}\cite{private_dmrg}  & -0.761826  & == & ==   \\
 {\rm iPEPS}\cite{private_tn} &   ==  & == &  -0.75333 \\
 {\rm iPEPS}$^*$\cite{stripe} &   ==  & == &  -0.767(2) \\
 {\rm This work} & -0.76276(5)  & -0.75867(3) & -0.75842(5) \\
 {\rm This work}$^*$ & -0.7654(1)  & -0.7618(1)   & -0.7605(1)\\
      \hline
  \end{tabular}
\caption{Best variational energies and extrapolated energies 
(methods marked by  $^*$) for various methods as compared with  VAFQMC (last  raws) for the $U/t=8$ Hubbard model at $\delta=1/8$. The numbers in parenthesis represent error bars or uncertainties in the extrapolations in the  last digit. Other quantum chemistry methods, using also multi reference (MR) expansion,  have be shown ''the need for a much larger MR expansion than that afforded"\cite{zhang_benchmark}.}
\label{tab_compare}
\end{table}

\begin{figure}[htbp]
  \centering
      \includegraphics[width=9cm]{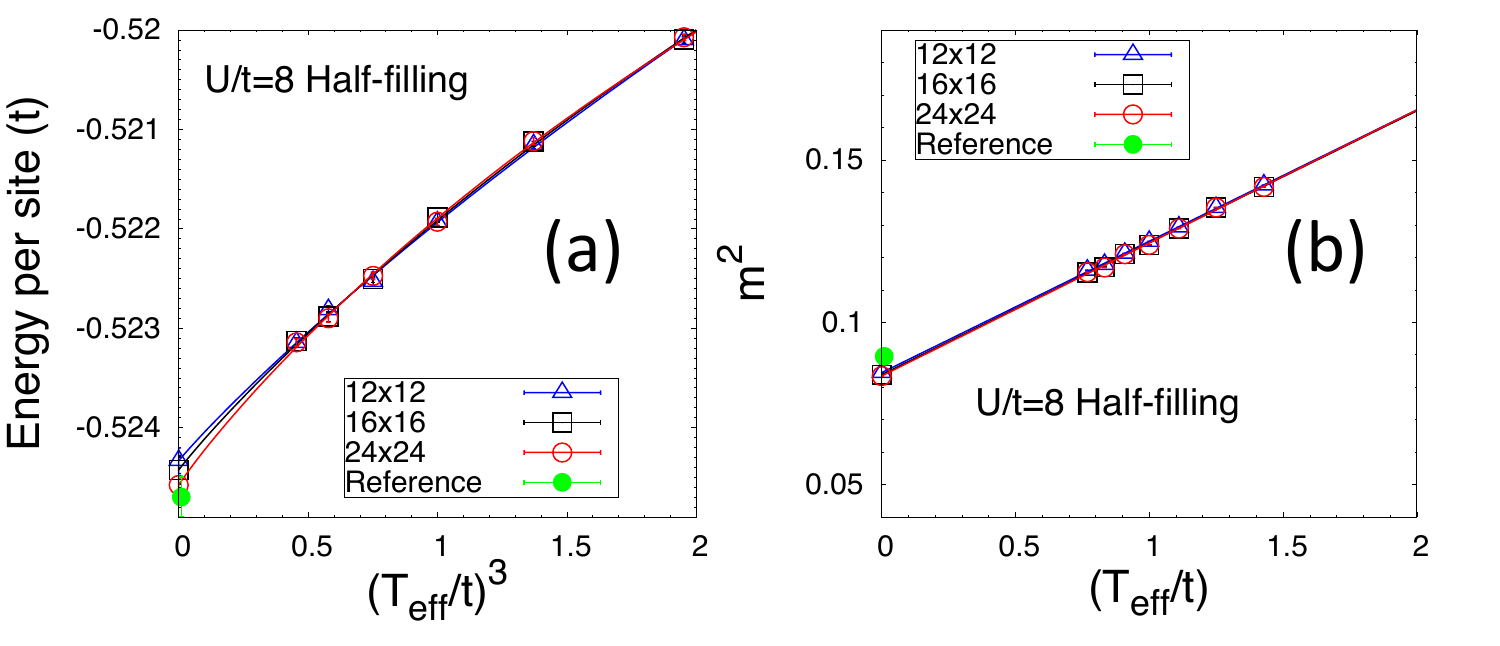}
           \caption{Energy per site (a) and spin-spin correlation function at maximum distance (b) converging to the square antiferromagnetic order parameter $m^2$  in the thermodynamic limit. All calculations 
	   refer to half-filling and are shown 
	                  as a function of the  effective temperature $T_{\mathrm{eff}} = 1/\tau$.
 Values at zero horizontal axis result from  extrapolations.
 The estimated exact reference values in the thermodynamic limit are taken from 
 Ref.\onlinecite{zhang_benchmark} for the energy and  Ref.\onlinecite{seki_benchmark} for $m$.}
        \label{half}
 \end{figure}

Finally the energy and magnetic order parameter are compared at half-filling. In this case there is no sign  problem  and in principle large imaginary times can  be employed without particular difficulties. However, 
$\tau t \le 1.3$ has been chosen also in this case,
 in order to show the strength of this  approach even when short time projections are employed.
For the  order parameter  finite effective temperature scaling analysis
implies convergence linear  in $T_{\mathrm{eff}}$ much slower than the energy (see App.\ref{sec:ftemp}). Nevertheless remarkably accurate 
results  can be obtained in both cases, as shown in Fig.~\ref{half}.
At finite effective temperature $T_{\mathrm{eff}}$ very weak finite  size effects are seen, showing once more the great advantage to use TABC with this approach.
The error in the $T_{\mathrm{eff}}\to 0$ extrapolated energy is then 
compatible with the reference one within  the error bars (this work 
extrapolation $E_h=-0.52443(14)$, reference $-0.5247(2)$) whereas the 
extrapolated order parameter ($0.2894(16)$)  is in very  good agreement with 
the benchmark one ($0.2991(2)$).

\section{The Hartree-Fock stripe phase}
\label{sec:hf}
Also within the Hartree-Fock, namely by considering a simple Slater determinant 
or also uncorrelated BCS mean-field wave functions, very few established 
results are known, if we allow also nonuniform 
solutions\cite{schulz_hf,zhang_hf}, i.e. non translation invariant solutions with large unit cell. At half filling it is clear that the simple antiferromagnetic solution, with two site per unit cell, is stable for any $U>0$ as it has been shown analytically by  Hirsch\cite{hirsch}. On the other hand it is well-known that no superconducting solution is possible for $U>0$.  
Restricting the  solution to  a small unit cell containing only two sites leads to a phase separation instability\cite{emery}, namely the  energy per hole:
\begin{equation} \label{ehole:eq}
e_h(\delta)= { \left(e(\delta)-e(0)\right) \over \delta}
\end{equation}
acquires a minimum at a finite doping $\delta_c$ and for any doping $\delta \le \delta_c$ it is more convenient to expel the holes from a pure antiferromagnet  with energy per site $e(0)$ in a region with appropriate size containing an hole rich phase at doping $\delta_c$. After this construction (see e.g. App.\ref{sec:conv}) $e_h(\delta)$ will be 
constant in the thermodynamic limit and equal to $e_h(\delta_c)$ for all dopings $\delta \le \delta_c$. This represents  
a fingerprint of phase separation and the study of the energy per hole can be also considered  an implicit  way to build a nonuniform solution (when  a minimum is found) within not only HF but any variational approach based on a translationally  invariant  ansatz.

\begin{figure}[htbp]
  \centering
      \includegraphics[width=9cm]{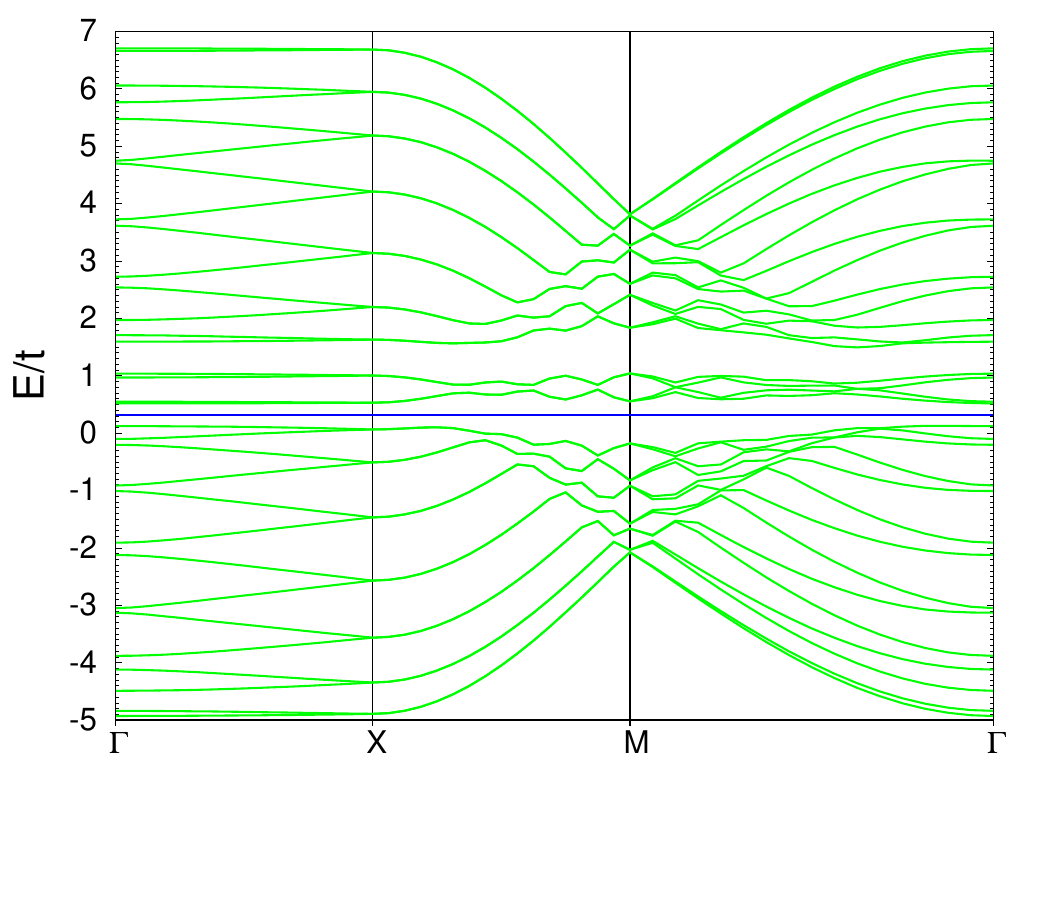}
           \caption{Hartree-Fock bands for the $W=8$ stripe at $U/t=8$ 
	   and doping $\delta=1/8$. There are 32 different bands and only the lowest energy $14$ bands are completely occupied at this filling, implying insulating behavior, as the Fermi energy (blue line) clearly separates the lowest occupied bands from the empty ones. Here $\Gamma=(0,0)$,$X=(\pi/16,0)$ and $M=(\pi/16,\pi/2)$.} 
        \label{hfbands}
 \end{figure}

However a better way to expel holes from an antiferromagnet is found by means of the stripe solution already reported in Fig.~\ref{stripe_4x16}.
In particular when the doping $\delta = 1/W$ the HF bands show a clear insulating behavior because the 
unit cell $2 \times 2 W$ contains $4W$ doubly-degenerate bands\footnote{Spin-up and spin-down bands are degenerate due to the reflection symmetry around any stripe, namely a  vertical  axis with maximum hole density, see Fig.~\ref{stripe_4x16}. This transformation  changes the spin direction of the antiferromagnetic  field, namely the spin-up with the spin-down HF Hamiltonian, that therefore have the same spectrum of eigenvalues.}  and $2W-2$ are fully occupied, as is the case for $\delta=1/8$ shown in Fig.~\ref{hfbands}.
The insulating  nature of this stripe solution was also pointed out in weak coupling HF theory by H. Schulz\cite{schulz_hf} who discovered first the  stability of incommensurate magnetic states with finite wavevector $Q=(\pi\pm \pi \delta,\pi)$ close to the antiferromagnetic one $Q=(\pi,\pi)$, and the opening of a full gap away from half filling within HF.
The occurrence of a finite gap in this case is easily  understood not only because 
the mean-field HF Hamiltonian turns out to have a gap, 
 but also for the following simple general 
argument holding also in the correlated case.
This incommensurate state  is adiabatically connected to the insulator
having equally  spaced empty (i.e.  with  no-electrons) vertical lines of sites 
separating half-filled antiferromagnetic insulating  regions.

\begin{figure}[htbp]
  \centering
      \includegraphics[width=9cm]{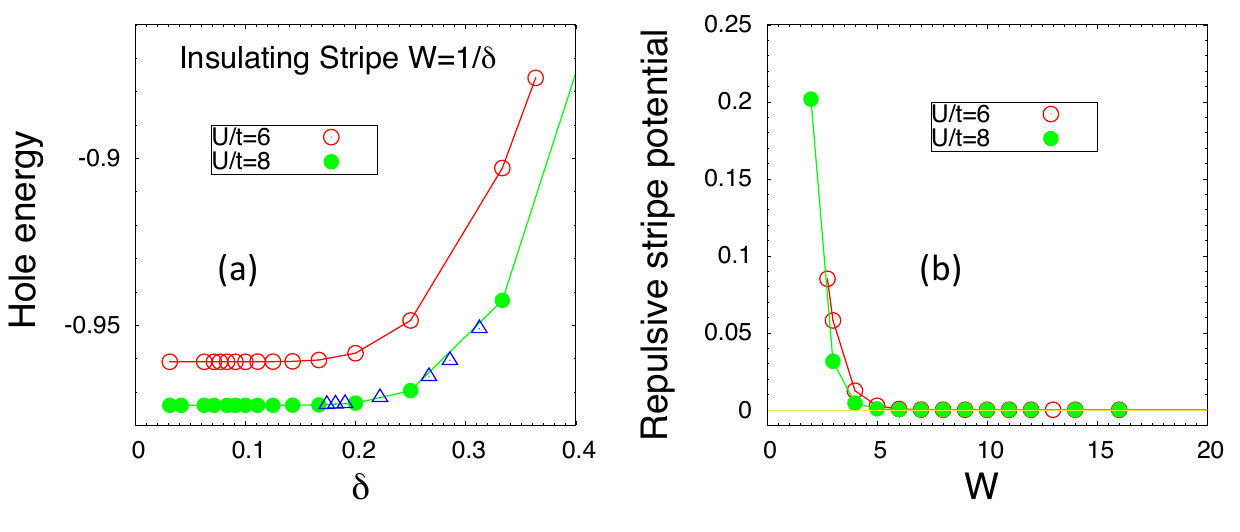}
           \caption{Hole energy (a) for the stripe of width $W$ commensurate with the doping $\delta$ within Hartree-Fock. The blue triangles for $U/t=8$ correspond to intermediate dopings obtained by applying the composition rule described at  the end of Sec.~\ref{sec:hf}. (b)   Effective repulsion energy between equally spaced  vertical stripes as a function of their distance $W$. } 
        \label{hf}
 \end{figure}

In this way the HF solution can avoid phase separation but the 
corresponding energy per hole is almost flat at small doping (see Fig.~\ref{hf}a) that  is almost 
equivalent to phase  separation.  Indeed at small doping the stripes  
are very far apart $W ={1\over \delta}$ and do not interact at all, as can be
appreciated in Fig.~\ref{hf}b where the interaction between two vertical  stripes at distance $W$ is given by $I(W)=e_h(\delta=1/W)-e_h(\delta=1/\infty)$. 
$I(W)$ here defined represents the energy cost  per hole for  two stripes being at finite distance $W$ rather than at infinite distance, where they do not interact.

At non commensurate doping, for instance  $1/8 < \delta < 1/7$ it is possible to  verify that such kind of insulating solutions at $\delta =1/7$ and 
$\delta =1/8$ can be joined together by forming a smooth doping dependent insulating stripe phase at any intermediate doping (see for instance Fig.\ref{hf}a for $U/t=8$). Many stripes at positions 
$r_{x_i}$ of the lattice can be thought  to interact by means of a pairwise repulsive interaction $\propto \sum\limits_{i<j} I(r_{x_i}-r_{x_j})$, with  interaction $I(W)$ as the one computed in Fig.~\ref{hf}b. Thus the rule for obtaining the minimum HF energy is as follows: {\em  in order to get the appropriate hole density $\delta$  one alternates energetically more expensive defect stripes of length $W$ smaller than $[1/\delta+1]$
($W=7$ in this case) by placing them as far as possible.}

\section{Results for the correlated phase diagram}
\label{sec:corrps}
The rule determined in the previous section to identify the minimum energy insulating  stripe solution has been verified also in the  correlated case up to  the maximum $\tau$ possible with essentially no sign problem, namely 
$\tau t=1.3,1.73,2.6$ for $U/t=8,6,4$, respectively.
Therefore  first the stripe wave function (breaking translation and spin symmetry) is computed at commensurate doping $\delta=1/W$, where $W$, being the distance between equally spaced stripes, is an integer. Then, in order to account for  intermediate dopings,  the corresponding energies are  interpolated. Moreover, in order to estimate the error to the exact $T_{\mathrm{eff}}=0$ limit,  the results are extrapolated to  this limit.
Hence the lowest  energy of the uniform solution is estimated, that is parametrized by the 
following mean-field Hamiltonian:
\begin{eqnarray}
H_{MF}(\bm{\alpha_0})  &=& \bar K - \mu_0 \hat N +\left[
\Delta_{AF} \sum_R (-1)^{r_x+r_y} c^{\dag}_{R,\uparrow} c_{R,\downarrow} \right. \nonumber \\
&+&\left. 2 \Delta_{x^2-y^2} \sum_k (\cos k_x - \cos k_y ) c^{\dag}_{k,\uparrow} c^{\dag}_{-k,\downarrow} + {\rm h.c.} \right] \nonumber  \\
\label{hmf}
\end{eqnarray}
where $\bar K=\sum\limits_{i,j,\sigma} t_{R,R^\prime} c^\dag_{R,\sigma}
c_{R^\prime,\sigma}$ is the translation invariant kinetic energy
defined here by the nearest and next nearest neighbor  hopping $t$ and
$t^\prime$, namely $t_{R,R^\prime}=-t$ ($t_{R,R^\prime}=-t^\prime$)
if $R$ and $R^\prime$ are (next) nearest neighbor sites, $\Delta_{x^2-y^2}$ 
determines  the gap function for a $d_{x^2-y^2}$ BCS superconductor, $\Delta_{AF}$ 
is the mean-field gap due to a commensurate antiferromagnetic state, 
and $\mu_0$ is the mean-field
chemical potential value.

By the proposed energy optimization method,
as the  doping is decreased  clear evidence of an 
instability towards $d_{x^2-y^2}$ BCS pairing is found, because as soon as 
$\Delta_{x^2-y^2}$ turns out to be non zero in the thermodynamic limit, 
 the variational wave function $|\psi_\tau\rangle$, breaks the $U(1)$ symmetry $c_{R,\sigma} \to \exp( i \theta) c_{R,\sigma}$ 
of the  Hamiltonian, acquiring the best possible energy for a  uniform 
superconducting phase.
In this approach $\Delta_{AF}$ turns out to be non zero only at half filling but it is important to emphasize that, despite the uncorrelated HF case, a non zero $d_{x^2-y^2}$ BCS pairing is possible at finite doping for $T_{\mathrm{eff}}$ lower than a critical effective temperature, and this represents one of the most important  effect of electron correlation, as it is not present in the infinite  $T_{\mathrm{eff}}$ HF theory.

\begin{figure*}[htbp]
  \centering
      \includegraphics[width=17cm]{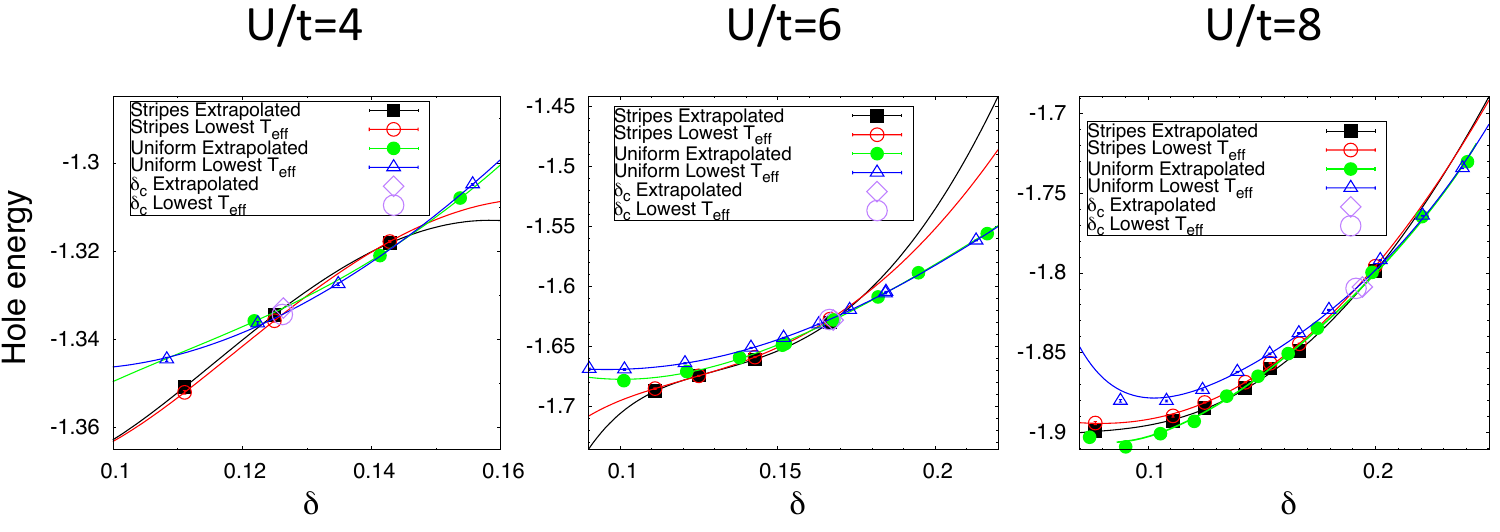}
           \caption{Determination of the critical doping where the transition between the uniform  phase and the stripe phase occurs with lower energy at small doping.  The empty big dots (tilted squares) represent the transition points corresponding to  the best variational estimates (extrapolated to zero temperature ones) at the lowest effective temperatures $T_{\mathrm{eff}}$} 
        \label{phase_stripe}
 \end{figure*}

In this way  we can determine the transition between the  translationally invariant  phase and the 
stripe phase as reported in Fig.~\ref{phase_stripe} for the three representative values of $U/t$. Notice also that the location of the transition points does not depend much on the extrapolation, clearly supporting that the phase diagram converges quite fast by lowering $T_{eff}$. 
 As in HF theory, away from the commensurate dopings $1/W$,  an  insulating 
 phase covering a continuum of different dopings can be constructed by appropriately joining such commensurate 
 solutions, i.e. a doping $2/13$ with insulating properties 
 can be easily obtained by alternating a $W=7$ stripe with a $W=6$ one. 
This is actually the case in  a $26\times 12$ 
cluster, within VAFQMC, namely the $6+7+6+7$ alternating stripe length solution has an energy  slightly lower than the corresponding 
one $6+6+7+7$, thus satisfying the HF rule stated at the end of the previous section.
This phase, competing with the uniform $d_{x^2-y^2}$ BCS one, has a full charge gap in $H_{MF}(\bm{\alpha_0})$, a property that cannot be changed by a small $\tau-$projection. Moreover  no coexistence of stripe and BCS order was found within VAFQMC. 

\begin{figure}[htbp]
  \centering
      \includegraphics[width=8.6cm]{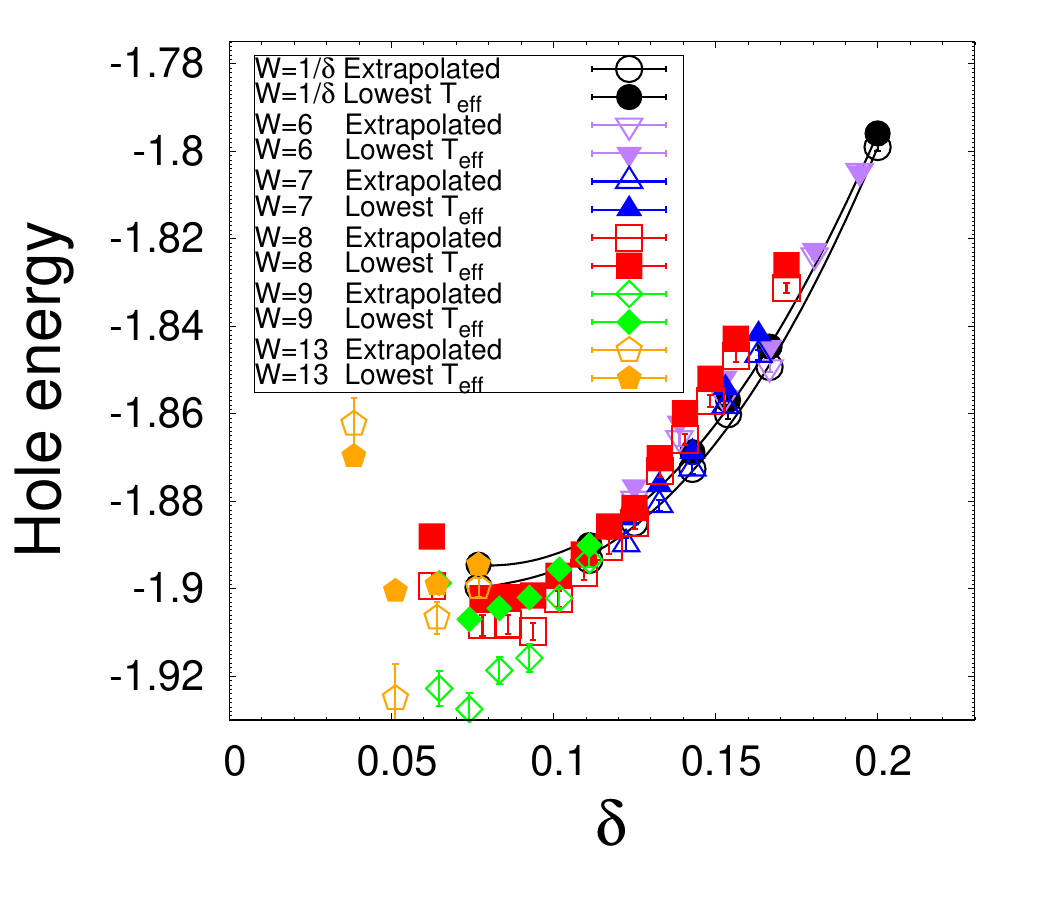}
       \caption{ Hole energy at $U/t=8$ for equally spaced stripes of different length $W$ as a function of the doping $\delta$, by considering clusters of dimension larger than $12 \times 2 W$, namely enough to have negligible finite size effects. For doping around $\delta\simeq {1\over 8}$ 
       stripes with length $W< {1 \over \delta}$ become energetically stable and acquire metallic character as the mean-field bands defining the variational 
       wave function used in this work are only partially filled.
The commensurate stripe results with $\delta =1/W$ are connected by black lines.}
       \label{stripevsW}
 \end{figure}

In summary, according to  
Fig.~\ref{phase_stripe}, it  is possible to conclude that  
an insulating stripe phase acquires an energy lower  than the 
corresponding  uniform phase already at quite large doping around $\delta \simeq 20\%$.
Nevertheless, it is important  to remark that the uniform phase 
has a very  good energy, competing with the lowest possible ones.
Indeed in Fig.~\ref{16rung} the uniform $d-wave$ phase  looses 
only  a tiny energy (less than $0.001t$ per site at the lowest $T_{\mathrm{eff}}$) 
vs the nonuniform stripe phases. Interestingly just around 
the doping $\delta =1/8$ equally spaced stripes with length $W< 1/\delta$ appear lowest in energy, as can be seen in Fig.~\ref{stripevsW}, where the energy of such metallic (at least within the mean-field Hamiltonian $H_{MF}(\bm{\alpha_0})$)  
stripes are compared with the commensurate insulating ones (black lines).
For instance the $W=9$ stripe at doping around $1/13$ (i.e. a  metallic stripe 
with $W<1/\delta$) has an hole energy more than   $0.01t$ below the corresponding insulating stripe with $W=13$, a huge gain that can be hardly explained by artifacts of the approximations (finite $T_{\mathrm{eff}}$ and/or finite clusters). 
In this plot we also see that phase separation (a minimum of the hole energy) 
cannot be excluded for doping $\delta \lesssim 5\%$ but the results strongly depend on the extrapolation and requires therefore more accuracy to be settled.
By judging from the extrapolated values it is plausible to expect a similar 
flat behavior of the  energy per hole at low doping as in the corresponding 
HF plot (Fig.\ref{hf}a). However, within HF, a metallic solution with $W < 1/\delta$ was never found to have the  lowest energy at low doping, again in agreement with Ref.\onlinecite{schulz_hf}.

\begin{figure}[htbp]
  \centering
      \includegraphics[width=9.6cm]{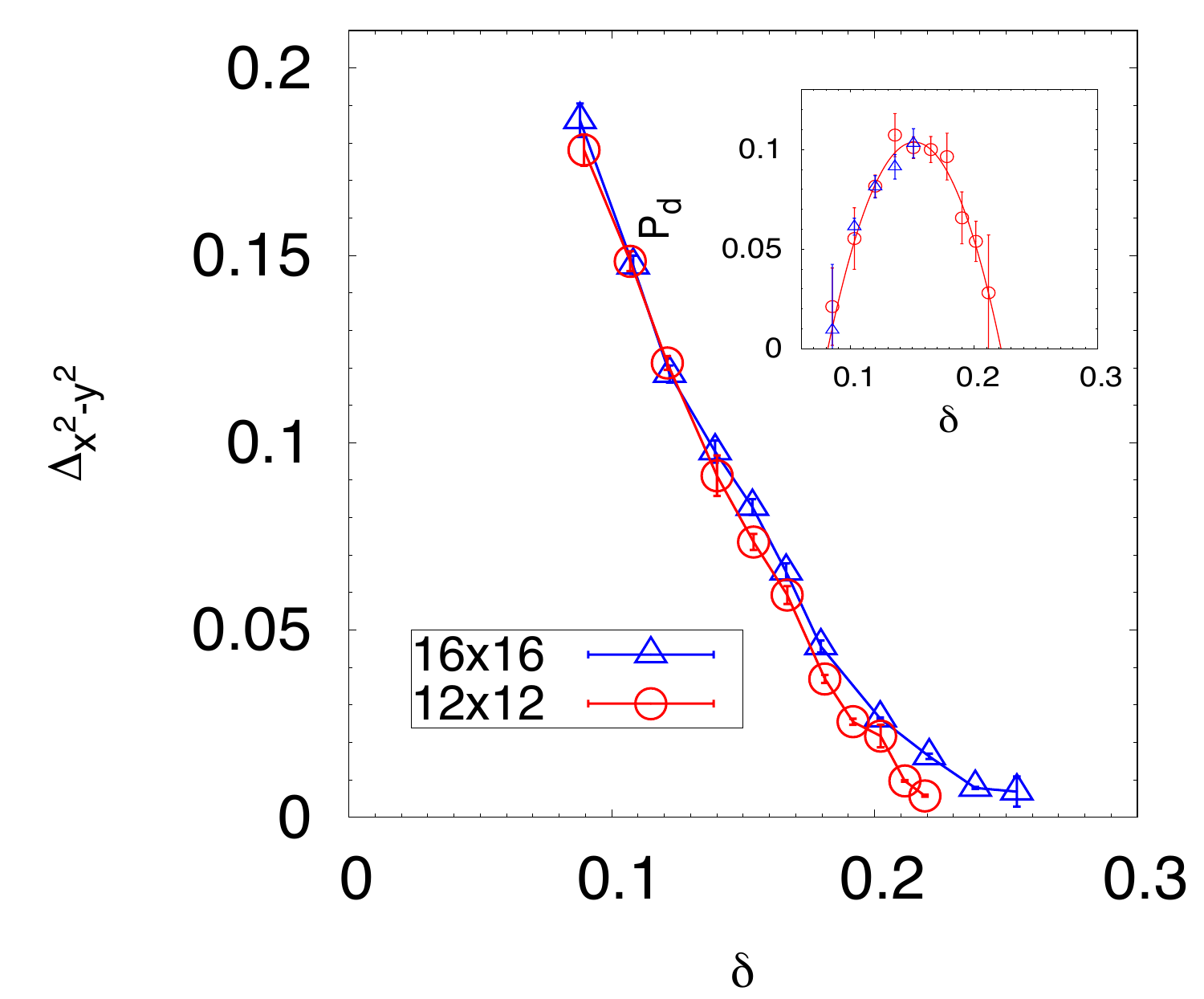}
              \caption{$\Delta_{x^2-y^2}$ BCS pairing variational parameter
	      in the  mean-field Hamiltonian $H_{MF}(\bm{\alpha_0})$ (see Eq.~\ref{hmf}) 
	      defining the spatially  uniform 
	      variational ansatz of Eq.~(\ref{wf}), as a function 
	       of the doping  $\delta$ for the lowest
	      $T_{\mathrm{eff}}$ and largest clusters considered. Inset: value of the corresponding BCS order parameter evaluated with  the $T_{\mathrm{eff}}\to 0$ linear  extrapolation of the pairing-paring correlations at distance $L/2$. The error bars include also uncertainty in this extrapolation.}
       \label{showdelta}
 \end{figure}

In the following it is worth to discuss rather extensively 
the main result of this work, namely that,
at large doping the stripe melts and  
a small but sizable superconducting d-wave order clearly remains within the present approach, in the sense that the energy is lowered in the thermodynamic limit by breaking the $U(1)$ global symmetry related to number of particle conservation. This is 
because both the mean-field Hamiltonians $H_{MF}(\bm{\alpha_0})$ and 
$H_{MF}(\bm{\alpha})$ clearly support  a broken symmetry solution of this type 
when $\Delta_{x^2-y^2}>0$ in Eq.~\ref{hmf}.  
Indeed,  in Fig.~\ref{showdelta}  the superconducting gap $\Delta_{x^2-y^2}$ 
parameter, corresponding to $H_{MF}(\bm{\alpha_0})$, is displayed 
for $U/t=8$ as a function of the doping.
When the  $\Delta_{x^2-y^2}$ is large there are negligible size effects.
However when it approaches zero, a long slowly decaying tail shows evident size effects. Indeed with the largest affordable sizes it is possible that the calculated critical doping $\delta_c$ above which a symmetric phase is stable indicates only a sharp crossover region 
separating a phase with sizable strong coupling superconductivity from another phase with very small order parameter, compatible 
with  an exponentially small 
pairing of the  Kohn-Luttinger type\cite{hlubina,prokovev}. 
Indeed, within weak coupling theory, the $d-wave$ superconducting phase should 
remain stable up to $\simeq 40\%$ doping, a doping much larger than the one detected in the present work. 

\begin{figure*}[htbp]
  \centering
      \includegraphics[width=18cm]{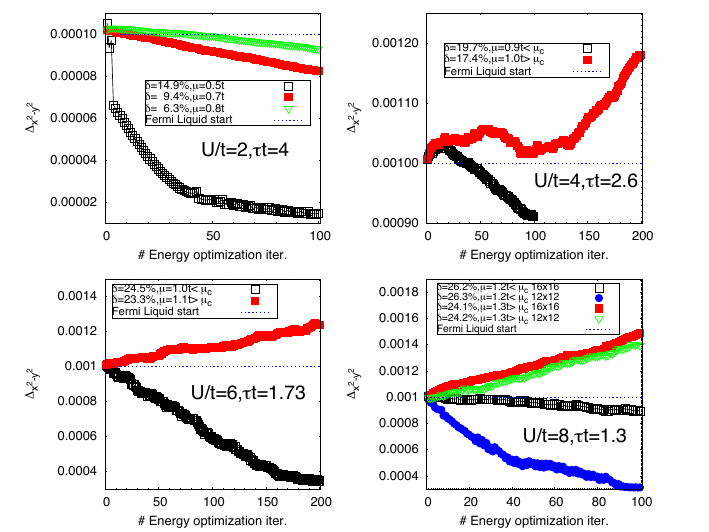}
              \caption{$\Delta_{x^2-y^2}$ BCS pairing variational parameter
	      in the  mean-field Hamiltonian $H_{MF}(\bm{\alpha_0})$ 
	      defining the spatially  uniform 
	      variational ansatz of Eq.~(\ref{wf}), as a function 
	       of the energy optimization iterations at the lowest 
	      $T_{\mathrm{eff}}$ (i.e. largest $\tau t=4,2.6,1.73,1.3$ for $U/t=2,4,6,8$, respectively) and $16\times 16$ cluster.  For all cases the initial wave function was prepared with $\Delta_{x^2-y^2}=0$ defining the Fermi liquid state. Then this parameter was set to a small value and the iterative stochastic optimization was started. In these calculations, in order to reduce at most the stochastic bias, at  each iteration a quite large number ($>3\times10^5$) of samples for evaluating energy derivatives was used, in a way that  they are sensitive to very small  energy differences  $< 10^{-5} t$. Optimization of the grand potential $\Omega$ at fixed chemical potential $\mu$ was employed in all cases shown. The corresponding doping was estimated by ${\partial \Omega \over \partial \mu} =-(1-\delta) N_s$ using the data for $\Omega$ reported in the supplementary material SI  extrapolated to the $T_{\mathrm{eff}}\to 0$ limit. These estimates are consistent with the direct evaluation of the doping $ \delta =1- {\langle \psi_\tau | \hat N | \psi_\tau \rangle \over N_s \langle \psi_\tau | \psi_\tau \rangle } $  at the lowest $T_{\mathrm{eff}}$ within half the confidence doping interval reported for each $U/t \ne 2$.}
       \label{pdvsiter}
 \end{figure*}

As shown in Fig.~\ref{pdvsiter}, within this approach, 
the critical doping  characterizing a strong coupling superconductivity (i.e. a non negligible pairing) can be 
identified because for $\mu \le \mu_c$ the energy 
clearly improves by increasing the value of $\Delta_{x^2-y^2}$ even when starting from a negligible value, representing the symmetric Fermi liquid ground state.
Nevertheless 
there may be non negligible size effects as shown for $U/t=8$, especially 
considering the difficulty to locate the end of a long tiny tail of the order parameter, as discussed before. Moreover,  for  smaller $U$ these size effects are expected even larger, but much larger cluster simulations are prohibitive at present. For this reason for $U/t= 2$ ($U/t=4$) a much larger mesh for the TABC was used with 
$64\times 64$ ($32\times 32$)  different twists in the BZ, instead of the smaller $16\times 16$ grid used for $U/t=8$ and $U/t=6$. In a mean-field calculation a $64\times 64$ TABC grid in a $16\times 16$ square lattice  corresponds to a $1024\times 1024$ cluster with PBC, large  enough to probe even a very tiny but non negligible gap parameter $\Delta_{x^2-y^2}$.
Remarkably, at $U/t=2$ no evidence of such a tiny value of the order parameter is found in all relevant doping region, as can be clearly  seen in  Fig.~\ref{pdvsiter}.

\begin{figure}[htbp]
  \centering
      \includegraphics[width=8.6cm]{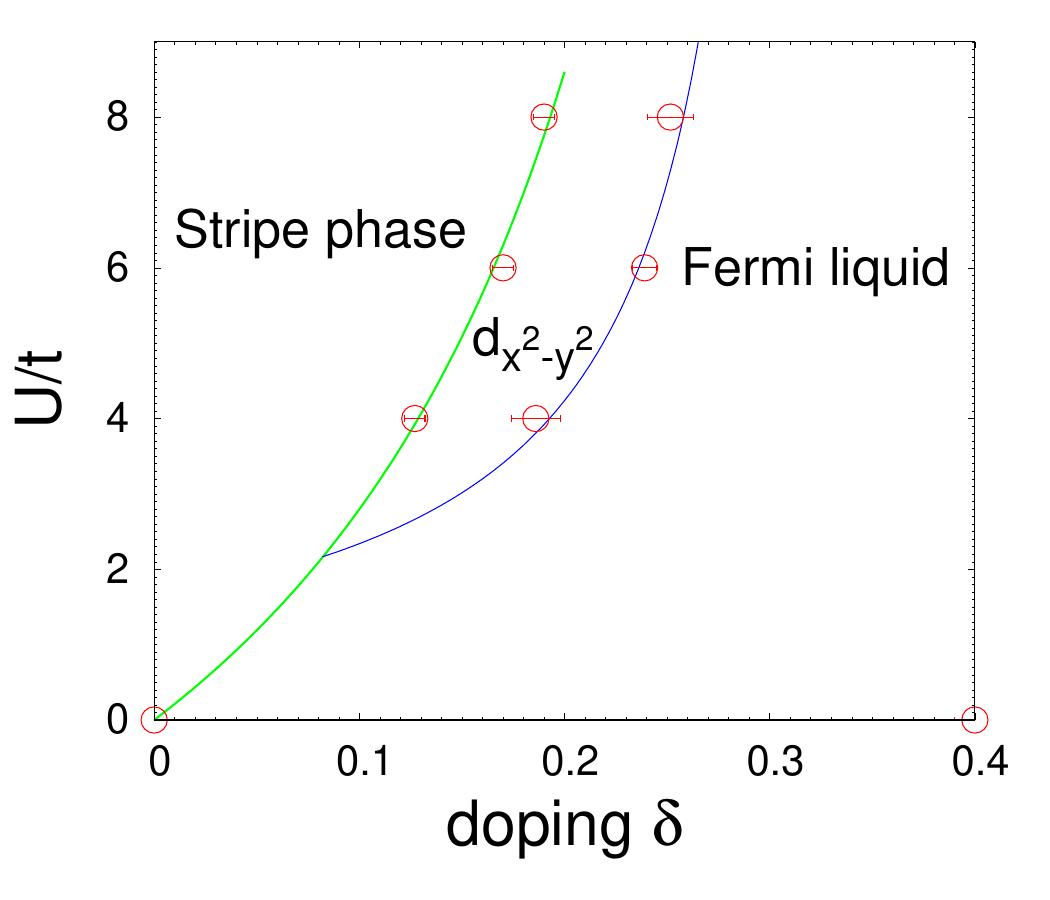}
       \caption{ Phase diagram of the Hubbard  model by VAFQMC.  
	 Blue and green continuous lines are simple interpolations of the transition points (open circles, $\delta=19\%-25.2\%,17\%-23.9\%,12.7\%-18.6\%$ for $U/t=8,6,4$, respectively). The blue line is also  determined by the condition that for $U/t=2$ no BCS pairing has been detected, while  the green line by the one   that the incommensurate spin order cannot show up at non zero doping for $U\to0$\cite{schulz_hf}.
       In the stripe phase there may be  a transition between an insulating stripe and  a small doping 
       metallic stripe, as suggested by Fig.~\ref{stripevsW} for $U/t=8$. In this transition 
       the mean-field VAFQMC Hamiltonian $H_{MF}(\bm{\alpha_0})$ 
               has a finite  gap (bands are completely filled) or a vanishing one, respectively.
       }

       \label{phase}
 \end{figure}

This small but relevant strong coupling region where d-wave 
superconductivity 
appears to be stable in the Hubbard model is in agreement with the  
old findings by cluster dynamical mean-field theory\cite{jarrell_pair}, though 
it is located at a slight higher doping, namely
at 10\% the stripe should be stable also for $U/t=4$. Moreover it is 
worth mentioning 
 a recent  
VMC calculation\cite{Imada} employed at $U/t=10$ and density 
matrix embedding theory(DMET)\cite{Chan} supporting also the stability of 
a $d-$wave superconductivity in a small doping region. On  the other hand 
the proposed phase diagram is also in agreement with the claim\cite{absence} of absence of $d-$wave superconductivity in the 2D Hubbard model, because in this work the authors  refer mainly to doping $1/8$, where no BCS pairing was found also in  this work.
The final phase diagram is therefore reported in 
 Fig.~\ref{phase}.

\section{Conclusions}
In this work the  VAFQMC method, able to exploit the power of the auxiliary field technique combined with the simplicity and generality of the standard variational  quantum Monte Carlo method, has been introduced. By combining these two successful approaches 
to strongly  correlated systems, it is  possible to estimate 
the evolution of the  phase diagram of a lattice model,
by improving systematically the accuracy of the correlation term, starting from the HF approximation. This is achieved by lowering an effective temperature $T_{\mathrm{eff}}$ up to a value that turns out to be low enough to provide very accurate variational energies of the Hubbard model in the thermodynamic limit, while remaining far from any sign problem instability, so far affecting most fermion quantum Monte Carlo techniques.
Within the assumption that the  tiny  energy gain  
between different broken symmetry phases of the model  can be sorted out 
by a careful optimization of a mean-field ansatz in presence of an accurate enough correlation term, the phase diagrams obtained in this way should be 
considered reliable, at least for the phases studied.

It is clear that the above assumption is very important  as, for instance
in the Hubbard model, we have estimated
an   energy per site error of about $2\times 10^{-3}t$
($4\times 10^{-4}t$) at $U/t=8$ ($U/t=4$), while  energy differences of the  various phases at the lowest $T_{\mathrm{eff}}$ considered can be even more than an order of magnitude smaller. This is not a difficulty of the present method, providing state of the art  variational energies (see supplementary information for detailed benchmark results), but clarifies once more the enormous challenge of  the electron correlation problem in numerical calculations. 
For instance on such clusters it is basically impossible to distinguish an unbroken  symmetric metallic phase  from a $d-$wave superconducting  one with small $\Delta_{x^2-y^2}$ gap parameter,  by judging merely on correlation functions (e.g. pairing correlations), and the above assumption provides a much more sensitive criterium for occurrence of broken symmetry phenomena. Relying on the above assumption, dramatically  helps numerical techniques, such as VAFQMC, able to work with boundary conditions fulfilling  the symmetries of $H$, that may be possibly broken in the thermodynamic limit.

In any case, the present work provides rigorous and accurate upper bounds  of the ground state energies  of the Hubbard model for several $U/t$ and doping values well converged in the thermodynamic limit (see supplementary information), that can be useful in future works on this important subject and/or for benchmarking new computational, theoretical or experimental techniques. The fact that extrapolated energies do not depend much on the ansatz adopted (all extrapolated energies agree within $0.001t$ per site) implies that energetic properties are settled within a reasonable accuracy. 
However, in order to get the right order without assuming it in the initial mean-field ansatz,  much larger projection times are required (see for instance Fig.\ref{16rung} where two completely different phases differ in energy by less than $0.001t$ per site), that are at present not possible.   

In this study many different mean-field solutions  have been attempted with several different initializations of parameters, including  
also broken time reversal 
solutions\cite{palee} and different modulations of stripes and pairing.
In all cases the smooth convergence of the energy as a function of $T_{\mathrm{eff}}$ has been verified. This is not only important for achieving reasonable extrapolations but especially for excluding being trapped in spurious local minima, that is a limitation of any non linear optimization technique.
The phase diagram presented here in Fig.\ref{phase} shows the phases so far determined with the  lowest possible variational energies among the chosen ansatzs. 


An important progress in this study is the control of finite size effects within VAFQMC.
The  conventional approach is to attempt exact calculations 
(usually very hard if not  impossible) 
at finite number of sites $N_s$ and try to  extrapolate  to ${1\over N_s}\to 0$. Within VAFQMC  it  is much simpler and indeed possible to extrapolate 
only in the effective temperature $T_{\mathrm{eff}}\to 0$ limit 
the finite $T_{\mathrm{eff}} \gtrsim t/2$ results.
To this purpose it is enough to perform $N_s\gtrsim 100$ calculations because they  are already very close to the  ${1\over N_s} \to 0$ limit, thanks to the very effective twisted average boundary condition method.

The results shown in this work are weakly affected by the sign 
problem. In general the average sign, appearing for instance in  the denominator of  Eq.~(\ref{defsign}), should decay exponentially to zero
both with $1/T_{\mathrm{eff}}$ and $N_s$, yielding prohibitive statistical errors.  However for $T_{\mathrm{eff}}$ large enough (i.e. the ones used here) the exponential decay in $N_s$  of the 
average sign is very  weak, due to a stability property of the adopted auxiliary field technique, that was  discovered several years ago\cite{sorella_young}. Hence  much larger clusters could be safely simulated, that is very promising considering also that  the ones reported here 
are not far from the state of the art simulations not vexed by the ''sign problem''.

In this work a large stripe phase in the Hubbard model exists as an 
effective way to expel the holes from a clean antiferromagnet. In this way genuine phase separation does not occur in the Hubbard model as a result of a very weak repulsive coupling between stripes at large distance. In practice, at  small doping, the energy per hole is almost constant, namely the inverse charge compressibility is vanishingly small,  in agreement with the phase separation scenario\cite{emery} and some previous numerical calculations\cite{bachelet,zhang_ps,mpbc}.

An important outcome of the present work is 
that a sizable superconducting phase is present in the Hubbard model with a non negligible order parameter only in the strong coupling regime.
For instance for $U/t=2$ no evidence of $d_{x^2-y^2}$ BCS order was 
probed and the maximum $\Delta_{x^2-y^2}\simeq 0.02t$ 
was obtained with the largest 
$U/t=8$ considered at about $20\%$ doping. 
The critical doping $\delta_c$ separating  the metal from the 
$d-$wave superconductor turns out to be much smaller than the one obtained with  weak coupling analysis\cite{hlubina,prokovev} and the question is how to reconcile the weak and the strong coupling limits.
In principle, as suggested by the results in  Fig.~\ref{showdelta}, 
a small tail with  very weak BCS pairing could distinguish a strong coupling phase with sizable pairing and a weak coupling phase with a sharp crossover (and not a transition) clearly separating  the two regimes.

The main features of cuprate superconductivity can be explained with the Hubbard model, but quantitatively there remain several unsolved issues. 
First, the stripes are completely filled by holes and do not match the half filled ones detected experimentally\cite{tranquada}. Second, the $d-$wave superconductivity is sizable but still far to explain $T_c\simeq 100K$ because the gap function, that is  growing when approaching the Mott insulating phase at zero doping (see Fig.\ref{showdelta}), is  limited by  the phase transition to the stripe phase. 

VAFQMC has been developed 
to study properties of strongly correlated systems 
in the thermodynamic limit. Obviously this method applies  also to 
finite systems for obtaining the best ground state estimates with  the largest possible projection time $\tau$. However for finite systems, the symmetry is never broken, and a  more accurate wave function should be used by restoring the symmetry with appropriate projection operators\cite{scuseria,imada_proj}, an improvement that can be done and should be worth doing for this particular purpose. 
Moreover, several new applications and extensions are 
possible within VAFQMC. 
The WF $|\psi_\tau\rangle$ can be further generalized to more 
realistic Hamiltonians, including  for instance long range  Coulomb  interaction or electron-phonon  coupling. 
The finite effective temperature $T_{\mathrm{eff}}$ here defined 
for a variational  wave function is the 
key to obtain very accurate  results well converged in the thermodynamic limit, a feature that could  
be possibly generalized to several other computational techniques, from 
tensor networks to machine learning variational wave functions.
\begin{acknowledgments}
\underbar{Acknowledgments:}
I am particularly in debt with A. Parola, N. Costa,  and  E. Tosatti for several 
suggestions for improving the  manuscript.
I am also tgrateful for very  useful discussions with G. Carleo, 
N. Costa, A. Tirelli,  F. Becca, L. Capriotti and  M.  Fabrizio. 
I acknowledge PRACE for awarding access to Marconi at CINECA Italy, and 
Riken collaboration for access  to HOKUSAI supercomputer  in Saitama Japan,
as well as  financial support from PRIN 2017BZPKSZ and the Simon's foundation.
\end{acknowledgments}
\appendix
\section{Finite effective temperature $T_{\mathrm{eff}}$ corrections for the infinite volume 
ground state projection}
\label{sec:ftemp}
Given a $\tau/2$-projected trial function $|\psi_\tau\rangle =\exp(-H \tau/2)
|\psi_{MF} \rangle $ the pseudo partition function,
\begin{equation}
 Z(\tau) = \langle \psi_\tau | \psi_\tau \rangle = \langle \psi_{MF} | \exp(-\tau H) | \psi_{MF}  \rangle,
\end{equation}
is considered in  the following, namely with an effective temperature $T_{\mathrm{eff}}={1 \over \tau}$ fixed when the infinite volume thermodynamic limit is employed.
Here the particular case of the Heisenberg model $H=J \sum\limits_{<ij>} \vec S_i \cdot \vec S_j $ and $|\psi_{MF} \rangle = |{\rm N\acute{e}el} \rangle $ is studied, but, due to 
universality, the results 
should apply to any model, where a classical symmetry, parametrized 
by a $M-$dimensional real vector is broken. In particular it  holds 
for the wave function 
of Eq.(1) in the main text, for collinear magnetic order $M=3$ (apply also for the stripe phase) 
and superconducting  $M=2$ one.
In the Heisenberg model, in $D$ dimensions,  the low energy limit can be studied  by introducing coherent states $\vec n$ with $ |\vec n|^2 =1$ and, by integrating over one of the two sublattices\cite{parola},  it follows that:
\begin{equation} \label{eqfund}
Z = \int d\left[ \vec n \right]_{0,\tau}   \exp\left\{ -{1 \over 2} \int\limits_{0}^\tau dt \int dr^D \left[  \chi | \partial_t \vec n |^2 + \Upsilon  \sum\limits_{\nu=1}^D |\partial_\nu \vec n|^2 \right] \right\}
\end{equation}
with the notations given in Ref.\onlinecite{fisher}, where $\Upsilon$ 
is the spin-wave stiffness and $\chi$ is the transverse susceptibility.
In the case of the pseudo partition function of Eq.(\ref{eqfund}) the following boundary conditions for the $M-$ component 
field $\vec n(\vec r,t)$ hold:
\begin{equation}
\vec n(\vec r,0) = \vec n(\vec r,\tau)= (1,0,\cdots,0)
\end{equation}
because the field at the boundary of the time interval 
is constrained to have the value of the trial function
$|\psi_{MF} \rangle $, a N\'eel state with the antiferromagnetic order along the x-axis.

Due to the above boundary conditions,  the $M-1$ component 
vector $\vec \Pi$, describing the field  fluctuations, and defined by:
\begin{equation} \label{defn}
\vec n(\vec r,t) = (1-{|\vec \Pi(\vec r,t)|^2\over 2},\vec \Pi(\vec r,t))+o(\Pi^2)
\end{equation}
acquires the following Fourier decomposition:
\begin{equation}
\vec \Pi(\vec r,t) =\sqrt{ 2 \over \tau V} \sum\limits_{n>0,\vec q} \vec \Pi_{n,
\vec q} \sin( \omega_n t) \exp ( i \vec q \cdot \vec r) 
\end{equation}
where $\vec q={ 2 \pi\over L} ( n_x,n_y,\cdots)$ are the momenta allowed by the periodic boundary conditions, whereas $\omega_n = { \pi n \over \tau}$ 
for integers $n>0$, in order to satisfy that the field vanishes at
$t=0$ and $t=\tau$. Notice that, in the usual partition function at finite temperature 
$T={ 1  \over \tau}$  the quantization of the frequencies $\omega_n= {  2 \pi n \over \tau}$ 
(also non positive integers allowed) is slightly different and the calculations can be easily 
generalized, but they have already been reported in previous 
works\cite{fisher,hasenfratz,hofmann}.
Therefore we just mention in the following  the standard finite temperature 
case, in order  to check the results derived in this section. 

With the above definition, the pseudo partition function acquires a Gaussian 
form at leading order, valid in the ordered phase:
\begin{eqnarray} 
Z&=&  \int \left[ d \vec \Pi \right] \exp\left\{ -{1 \over 2} \sum\limits_{n,\vec q} (\chi \omega_n^2 + \Upsilon |\vec q|^2 ) | \vec \Pi_{n,\vec q} |^2 \right\}  \nonumber \\ 
&\propto&  \exp\{ -{M-1 \over 2 } \sum\limits_{n,\vec q} \ln ( \chi \omega_n^2 + \Upsilon |\vec q|^2) \} \label{zeq}
\end{eqnarray}
On the other hand the average propagator over these Gaussian fluctuations 
can be readily evaluated:
\begin{equation} \label{propagator}
\langle  \Pi^\nu_{n,\vec q} \Pi^\nu_{n,-\vec q} \rangle = { 1 \over \chi \omega_n^2+ \Upsilon |\vec q|^2 }  
\end{equation}
where $\nu =1,2,\cdots M-1$ labels the components of the vector $\vec \Pi.$
\subsection{Energy and order parameter}
The expectation value of the energy $E(\tau)$ 
over the state $|\psi_\tau \rangle = 
\exp(- H \tau /2 ) | \psi_{MF} \rangle$ can be written as:
\begin{equation}
E(\tau) = - \partial_\tau \ln Z 
\end{equation}
Since $E(\tau)$  in  the expression (\ref{zeq}), depends on $\tau$ only by means of  $\omega_n$,   it follows that:
\begin{equation} \label{energy}
E(\tau) = -{(M-1)\over \tau}   \sum_{n>0,\vec q} { \chi \omega_n^2 \over 
\chi \omega_n^2 + \Upsilon |\vec q|^2 } 
\end{equation}
On the other hand the local magnetization:
\begin{equation}
 m(\tau) ={\langle \psi_\tau | { 1 \over V } \int d\vec r^D  (1 -{\Pi(\vec r,\tau/2)^2 \over 2})| \psi_\tau \rangle \over \langle \psi_\tau |  \psi_\tau \rangle }  
\end{equation}
can be easily evaluated by the knowledge of the propagator in Eq.(\ref{propagator}) and the definition of the component in the ordered x-direction in Eq.(\ref{defn}):
\begin{equation} \label{mag}
m(\tau) =1 -{  (M-1) \over V \tau} \sum\limits_{n>0,\vec q} { \sin({\omega_n \tau \over 2})^2  \over \chi \omega_n^2 + \Upsilon |\vec q|^2 }
\end{equation}
\subsection{Evaluation}
As in any field theory the above expressions diverge without introducing infinite counter terms. 
In the expression of the energy one can subtract and add one in the 
integrand, whereas in the correction of the magnetic moment it is enough to deal with particular care the $q=0$ mode (this mode may be also avoided by using twisted boundary conditions):
\begin{eqnarray}
 E/V  &=&  { M-1 \over \tau V} \sum\limits_{n>0,\vec q} { \Upsilon |\vec q|^2  \over  \chi \omega_n^2 + \Upsilon |\vec q|^2 } -{ M-1 \over \tau V } 
\sum\limits_{n>0,\vec q} 1  \nonumber \\ 
m (\tau) &=& 1 -{  (M-1) \over V \tau} \sum\limits_{n>0,\vec q } { \sin({\omega_n \tau \over 2})^2  \over \chi \omega_n^2 + \Upsilon |\vec q|^2 }
 \label{eqq0done}
\end{eqnarray}
The infinite RHS term  in the energy 
corresponds to an infinite shift of the ground state energy, a typical feature of  quantum 
field theories, 
as this shift becomes finite by introducing a cutoff, implicitly 
present in any physical model. 

In order to evaluate the above expressions it is useful to consider the 
following sum:
\begin{equation}
\sum_n   { 1 \over z + n } =\pi \cot(\pi z) 
\end{equation}
that is valid for any complex number $z$. In particular for 
$z={1 \over 2 } + i y$ we obtain:
\begin{equation} \label{form}
\sum_n { 1 \over {1 \over 2 } + i y + n } = -i \pi \tanh( \pi y) 
\end{equation}
whereas for $z=i y$ we get:
\begin{equation} \label{fore}
\sum_n { 1 \over  i y + n }  = -i \pi \coth(\pi y)
\end{equation}
Defining the velocity of the Goldstone's modes $c=\sqrt{\Upsilon \over \chi }$,
the sum over $n$ is extended to $-\infty \le n \le \infty$, 
using  the even dependence  on $n$ and taking into account the extra 
$n=0$ contribution  in the definition of $\epsilon_0$.
In this way, by using Eq.~\ref{fore} for evaluating the infinite sums the universal finite size corrections are given by:
\begin{eqnarray}
 E /V &=& { M-1 \over  2 \tau V } \sum\limits_{n=-\infty}^{\infty} \sum\limits_{\vec q} {  i  \sqrt{ \Upsilon} |q| \over 2} \left[ { 1 \over \sqrt{\chi} \omega_n + i \sqrt{\Upsilon} |q|} \right. \nonumber \\
&-& \left. { 1 \over \sqrt{\chi} \omega_n - i \sqrt{\Upsilon} |q|} \right] 
 - { M-1 \over 2 \tau V } \sum\limits_{n=-\infty}^{\infty}\sum\limits_{\vec q}  1  \nonumber \\
&=& \epsilon_0+ { ( M -1)  c  \over 2  V }  \sum\limits_{\vec q} |q|  (\coth ( c |q| \tau)-1) 
  \label{defelast}
\end{eqnarray}
where $\epsilon_0$ is the ''infinite'' ground state  energy 
corrected for the zero point harmonic energy  $\epsilon_0 = {M-1 \over 2 V} 
\sum\limits_{\vec q} c|q| - { M-1 \over \tau 2 V } \sum\limits_{n=-\infty}^{\infty}
\sum\limits_{\vec q} 1 $, a divergent expression depending on the cutoff but not on  the boundary conditions (i.e. it is the same divergent term appearing in Ref.\onlinecite{fisher}, by using the trace).
In the thermodynamic limit and fixed $\tau$ 
 the convergent contribution can be worked out, by replacing  sums over momenta with integrals (i.e.  ${1 \over V} \sum\limits_q \to \int ({dq\over 2 \pi})^D$), and changing the momentum scale $ \vec q \to c \tau \vec q$, so that:
\begin{equation}
 E/V =  \epsilon_0 +{(M-1) \over 2c^D}  f_e T_{\mathrm{eff}}^{D+1}
\end{equation}
where $f_e$ is a dimensional dependent constant defined by:
\begin{equation}
f_e = \int ({ dq\over 2\pi})^D   |q| (\coth( |q|)-1)
\end{equation}
that is clearly a convergent integral for any dimension $D\ge 1$ because
$\coth(|q|)$ converges to one exponentially for large $|q|$.
The values of $f_e$ are therefore $\pi\over 12$, $\zeta(3)\over 4 \pi$ ($\zeta(3)=1.20205690316$) and $\pi^2\over 240$ 
in $D=1$, $D=2$ and $D=3$, respectively.
In $D=1$ the  coefficient $f_e$ has not to be  taken for granted, since no true long range order 
is possible (see later), but the  qualitative behavior of the correction $\simeq T_{\mathrm{eff}}^2$ 
should hold as it is consistent  with  the corresponding  finite temperature correction derived in  
$D=1$ for the 1D-Hubbard model by means of the  Bethe Ansatz\cite{HubT}.

Thus, in the thermodynamic limit, the finite $\tau$ 
corrections to the energy depend only 
on the universal constant $f_e$, the number of components $M$ of the order parameter,  and  the velocity $c$ of the gapless Goldstone modes.
In D=2 the finite temperature corrections to the internal energy 
at infinite volume have been determined  
in Ref.\cite{hasenfratz}. For this purpose we have repeated the same type of calculation and 
get that, taking  into account the different boundary condition for the field (PBC in time 
for the trace $\tau=\beta=1/T$):
\begin{eqnarray}
E(T)/V - \epsilon_0 &=&  { ( M -1)  c  \over 2  }  \int ({dq\over 2 \pi})^D |q|  (\coth ( {c |q| \beta\over 2} )-1) \nonumber \\
 &=&  { (M-1) c \over 2 ({c\over 2 T})^{D+1} } f_e 
\end{eqnarray}
that is consistent with the previous calculation reported in D=2, i.e.\cite{hasenfratz}:
\begin{equation}
E(T)/V-\epsilon_0=  (M-1)  {\zeta(3) \over c^2\pi}  T^3 
\end{equation}

On the other hand, for the magnetic order parameter, 
by noticing that $\sin({\omega_n \tau \over 2} )^2=1$ 
for $n=2 \nu +1$ and zero otherwise,  the sum has to be evaluated for odd integer  frequencies $\omega_\nu= {\pi (2 \nu+1) \over \tau}$. 
For this purpose the sum 
$ \sum\limits_{ \nu=0 }^\infty { 1 \over \chi \omega_\nu^2 + \Upsilon |\vec q|^2} $ 
is extended to $ {1 \over 2 } \sum\limits_{\nu=-\infty}^\infty { 1 \over \chi \omega_\nu^2 + \Upsilon |\vec q|^2} $ because $\omega_{\nu} =-\omega_{-(\nu+1)}$ for $\nu< 0$. Therefore by applying  Eq.~\ref{form} 
for evaluating the following infinite sums, one  obtains:
\begin{eqnarray}
\Delta m (\tau) &=& - { M-1 \over 2V \tau } \sum\limits_{\nu, \vec q } 
  { i \over 2 \sqrt{\Upsilon} |q| } \left[ { 1 \over { \sqrt{\chi} \pi \over \tau } ( 2 \nu +1) +i \sqrt{ \Upsilon } |q| } \right. \nonumber \\
  &-& \left. 
{ 1 \over { \sqrt{\chi} \pi \over \tau } ( 2 \nu +1) -i \sqrt{ \Upsilon } |q| } \right]  \nonumber \\
  &=& -{ M-1 \over  4 V \sqrt{\chi \Upsilon} } \sum\limits_{\vec q \ne 0 } { (\tanh ( c |q| \tau/ 2 )-1 ) \over |q|} -\Delta m_0 \nonumber \\
  &=& -\Delta m_0 -{ M-1 \over  4 \sqrt{\chi \Upsilon} c^{D-1} } f_m  T_{\mathrm{eff}}^{D-1} \nonumber \\
  & & 
\end{eqnarray}
\noindent where $\Delta m_0 = {(M-1) \tau \over 8 V \chi}+  { M-1 \over 4 \sqrt{\chi \Upsilon}  V} \sum\limits_{\vec q \ne 0}  { 1 \over |q|} $ (the first term in $\Delta m_0$ is 
the $q=0$ contribution in Eq.~\ref{eqq0done} that is convergent for finite $\tau$) and 
the dimensional dependent constant  $f_m$ is defined by:
\begin{equation}
f_m = \int ({dq\over 2\pi})^D (\tanh( {|q| \over 2} )-1)/|q|  
\end{equation}
that converges in $D>1$.
The values of $f_m$ are therefore $-{\ln(2)\over \pi}$  and $-{1\over 12}$ for $D=2$ and 
$D=3$, respectively.

Thus, in the thermodynamic limit the finite $T_{\mathrm{eff}}$ 
corrections to the antiferromagnetic order parameter depend on 
the universal constant $f_m$, the number of components $M$ of the order parameter,  the velocity $c$ and the stiffness $\Upsilon$ of the gapless Goldstone modes.

Notice that the expression above for $f_m$  converges in $D=2$. 
Instead if periodic boundary conditions in $\tau$  are assumed for the field $\vec \Pi(x,t)$, corresponding to the standard finite temperature calculation at $\beta = \tau$, by repeating the same steps, it  follows that:
$\Delta m(\tau) = -{ M-1 \over 2 V \sqrt{\chi \Upsilon}} \sum\limits_{q \ne 0} {\coth(  c |q| \tau /2 ) \over |q| } $
which blows up for $D\le 2$, thus recovering  the Mermin-Wagner theorem\cite{Mermin}: 
no finite $m$ is possible at finite temperature for $D\le 2$.
The advantage of using the projection technique is therefore evident especially in  
$D=2$ for the study of broken symmetry phases, that are possible, 
for continuous symmetries, only at zero temperature.

When using  finite cylinders with  even $L_y$ all the above  results obtained for $D=1$ apply
because the $y$ momentum $q_y= {2 \pi n_y \over L_y }$ is quantized with  $n_y=0,1,\cdots,L_y-1$ 
and only the $q_y=0$ momentum value provides power law corrections in $1/\tau$ to energy 
and correlation functions (if converging). 
This is because all the other contributions acquire a finite
$\sim { 1\over L_y}$ gap and converge much faster.

\section{Convexity of the energy as a function density} 
\label{sec:conv}
One can divide a large system 
$A+B$ in two regions $A$ and $B$ containing a macroscopic number of sites $L_A$ and $L_B$ and electrons $N_A$ and $N_B$, respectively.
Then, by defining $E_{A+B}$ the energy of the total system, $E_A$ and $E_B$ the ones of the corresponding parts,  $\rho= {N_A+N_B\over L_A+L_B}=p \rho_1+(1-p) \rho_2$, $\rho_1={N_A\over L_A}$, $\rho_2= { N_B  \over L_B}$,  $p={L_A \over L_A+L_B}$, $e(\rho_1)=E_A/L_A$ and  $e(\rho_2)=E_B/L_B$ converged in  the  thermodynamic limit because the two parts are macroscopic,
then  $e(\rho) = {E_{A+B} \over L_A+L_B}\le { E_A+E_B \over L_A+L_B} =p e(\rho_1)+ (1-p) e(\rho_2)$ follows because one can neglect in the thermodynamic limit the surface term contribution to the energy separating the region A from the region B. This is  because, by assumption,  the model is short range as the hopping term connects only nearest neighbor sites and the interaction $U$ is on site.
The final inequality:
\begin{equation} \label{sp:ineq}
 e(\rho) \le p e(\rho_1)+ (1-p) e(\rho_2)
\end{equation}
therefore holds for arbitrary densities $\rho_1 \le \rho \le \rho_2$ and implies, the convexity property of the function $e(\rho)$ in the thermodynamic limit. 
Whenever phase separation occurs between two densities $\rho_1$ and $\rho_2$ the inequality~\ref{sp:ineq} turns to a strict equality, namely $e(\rho)$ is a linear function of the density because $p={ \rho-\rho_2 \over \rho_1-\rho_2}$, so that in particular the energy per hole in Eq.~\ref{ehole:eq} is constant for $\delta \le \delta_c$.

\bibliographystyle{apsrev4-1}
\bibliography{references.bib}
\end{document}


\title{Supplementary Material for: Stripes and $d-$wave superconductivity in the Hubbard model
by Variational Auxiliary Field quantum Monte Carlo }
\author{Sandro Sorella} 
\affiliation{International School for Advanced Studies (SISSA),
Via Bonomea 265, 34136, Trieste, Italy}
\affiliation{
Computational Materials Science Research Team, RIKEN Center for Computational Science (R-CCS), Kobe, Hyogo 650-0047, Japan}

\maketitle

\widetext
This supplementary material contains energies and correlation functions 
of the Hubbard model
obtained with the VAFQMC method described in the main text 
for various $L_x\times L_y$ cluster sizes. 
In all the tables error bars are between brackets with usual conventions.
Extrapolations to the $T_{\mathrm{eff}}=1/\tau\to0$ limit are obtained by using 
low order polynomials in $T_{\mathrm{eff}}$. In order to estimate the error 
bars in the extrapolated values a resampling technique is adopted, namely 
the fit is repeated several times $\simeq 1000$ 
using '' resampled'' data consistent with their 
error bars. This is possible because each data is obtained by an independent 
calculation and therefore there is no correlation between data corresponding to 
different $\tau$. Thus a resampled data can be considered a result of a different calculation compatible with  the error bar calculated with the original VAFQMC 
simulation. Then the error bars in the extrapolated quantities are obtained by estimating the corresponding standard deviations within the 
generated resampled data. 

As discussed in the main text, all the following  calculations 
have been  done by using twisted 
average boundary conditions, namely by imposing opposite twists 
for opposite spins:
\begin{eqnarray}
c_{r_x+L_x,r_y,\uparrow}&=&c_{r_x,r_y,\uparrow}\exp( i 2\pi  \theta_x ) \nonumber \\
c_{r_x,r_y+L_y,\uparrow}&=&c_{r_x,r_y,\uparrow}\exp( i 2\pi  \theta_y ) \nonumber \\
c_{r_x+L_x,r_y,\downarrow}&=&c_{r_x,r_y,\downarrow}\exp( -i 2\pi  \theta_x ) \nonumber \\
c_{r_x,r_y+L_y,\downarrow}&=&c_{r_x,r_y,\downarrow}\exp(-i 2\pi  \theta_y ) \nonumber \\
\end{eqnarray}
with $\theta_x=-1/2 + (i-1/2)/N_T$ and $\theta_y=-1/2+(j-1/2)/N_T$ 
for integers $1\le i,j \le N_T$.
Then, all the results  are averaged,  at fixed variational parameters, over the 
 $N_T^2=256$, ($N_T^2=1024$)  boundary conditions for $U/t=8,6$ ($U/t=4$). 

\section{ Translation invariant (uniform) parameterization}
In all the forthcoming results
the trial function used for the uniform phase 
is defined by means of the ground state
$\psi_{MF}$ of the following mean field Hamiltonian:
\begin{equation}
H_{MF}(\bm{\alpha})  = \bar K - \mu_0 N +\left[
\Delta_{AF} \sum_R (-1)^{x+y} c^{\dag}_{R,\uparrow} c_{R,\downarrow} +2 \Delta_{BCS}^{x^2-y^2} \sum_k (\cos k_x - \cos k_y ) c^{\dag}_{k,\uparrow} c^{\dag}_{-k,\downarrow} + {\rm h.c.} \right]
\end{equation}
where $\bar K=\sum\limits_{i,j,\sigma} t_{R,R^\prime} c^\dag_{R,\sigma} 
c_{R^\prime,\sigma}$ is the translation invariant kinetic energy 
parameterized by the nearest and next nearest neighbor  hopping $t$ and 
$t^\prime$, namely $t_{R,R^\prime}=-t$ ($t_{R,R^\prime}=-t^\prime$)  
if $R$ and $R^\prime$ are (next) nearest neighbor sites, 
and $\mu_0\ne 0$ is the mean-field  
chemical potential value, $N=\sum\limits_{R,\sigma}  c^{\dag}_{R,\sigma} c_{R,\sigma}$ being the
 total number of particle operator.
  $R=(x,y)$ is a lattice point belonging to the $lx \times ly$  
  cluster  lattice, namely
   $R=(x,y)$  with integers  $x$  and $y$, such that $ 1\le x \le L_x$ and $1\le y \le L_y$.  

In the half-filled case  $t^\prime=\mu_0=0$ is set and only  two parameters are 
optimized 
$\Delta_{AF}$ and $\Delta_{BCS}^{x^2-y^2}$. Conversely,  away from half-filling, 
(three parameters) 
$t^\prime$ and $\mu_0$ are turned on and $\Delta_{AF}$ is switched off 
by setting  it  to zero.

\subsection*{Supplementary Tables}

\begin{table}[h]
  \begin{tabular}{|l|l|l|l|l|l|l|}
      \hline
	 $\tau t$ &\multicolumn{3}{c|}{$E/N_s$} & \multicolumn{3}{c|}{$m^2$} \\   
      \hline
  $N_s$   & $12\times12$ & $16\times 16$ & $24\times24$    & $12\times12$ & $16\times 16$ & $24\times24$ \\ 
 \hline
 .7   & -0.518587(29)  & -0.518716(34) & -0.518712(28) & 0.142190(24)   & 0.141578(21)  & 0.141641(23)\\
 .8   & -0.520102(32)  & -0.520096(40) & -0.520075(33) & 0.135237(28)   & 0.135346(25)  & 0.135318(26) \\ 
 .9   & -0.521158(36)  & -0.521125(43) & -0.521120(35) & 0.129239(34)   & 0.128847(30)  & 0.128815(31) \\
 1.0  & -0.521936(36)  & -0.521887(45) & -0.521931(35) & 0.124827(39)   & 0.123821(33)  & 0.123818(35) \\
 1.1  & -0.522535(38)  & -0.522508(32) & -0.522475(37) & 0.121171(45)   & 0.121022(27)  & 0.120914(40) \\
 1.2  & -0.522817(40)  & -0.522884(48) & -0.522902(39) & 0.117734(62)   & 0.116919(40)  & 0.116849(46) \\
 1.3  & -0.523148(40)  & -0.523132(38) & -0.523143(38) & 0.115964(56)   & 0.115318(36)  & 0.115369(46) \\
 $\infty$ & -0.52434(13) & -0.52443(14) & -0.52458(7) & 0.08440(8) &  0.08377(6)  &  0.08339(7) \\
      \hline
 $\chi^2$ & 1.5 &  1.3   & 1.1 &   38 &  379 & 298.5   \\
      \hline
    \end{tabular}
  \caption{Energy per site  $E/N_s$  (spin spin correlation function $ {\langle \psi_\tau} | \vec S_R \cdot \vec S_{R^\prime} | \psi_{\tau} \rangle\over \langle \psi_{\tau}| \psi_{\tau}\rangle $  at the maximum distance, indicated here with  $m^2$) 
  of the half filled Hubbard model at $U/t=8$ as a   function of $\tau$ ($T_{\mathrm{eff}}={1\over \tau}$) extrapolated to the $\tau\to \infty$ ($T_{\mathrm{eff}}\to 0$) 
with a simple polynomial fit $E_0/N_s=E(\infty)+ A T_{\mathrm{eff}}^3+ B T_{\mathrm{eff}}^4$ ($m^2= m^2(\infty) + C T_{\mathrm{eff}}$)				        limit for various clusters. 
					The last raw indicates the reduced $\chi^2$  of the  fit. It  is large for the $m^2$ case because these numbers are much more sensitive to the  accuracy of the optimization.  
					    The numbers in parenthesis represent error bars in the  last digits.}
  \label{tab1}
  \end{table}

\begin{table}[h]
  \begin{tabular}{|l|l|l|l|l|l|l|l|l|}
    \hline
    $\tau t$ &  $\mu=1.2t$ & $\mu=1.3t$ & $\mu=1.4t$ & $\mu=1.5t$  & $\mu=1.6t$ & $\mu=1.65t$ & $\mu=1.7t$  \\
    \hline
 .8  	& -1.848336(25)	& -1.923309(26)	& -1.999942(26)	& -2.078202(37)	& -2.158311(37)	& -2.199297(40)	& -2.240705(35)	 \\
 .9  	& -1.849281(26)	& -1.924432(25)	& -2.001133(26)	& -2.079695(36)	& -2.160035(38)	& -2.200899(41)	& -2.242516(36)	 \\
 1.0  	& -1.850222(27)	& -1.925305(27)	& -2.002109(26)	& -2.080572(40)	& -2.161115(38)	& -2.202214(42)	& -2.243861(39)	 \\
 1.1  	& -1.850672(28)	& -1.925845(28)	& -2.002658(29)	& -2.081424(40)	& -2.162082(30)	& -2.203225(47)	& -2.244920(42)	 \\
 1.2  	& -1.851020(30)	& -1.926216(32)	& -2.003182(33)	& -2.081879(45)	& -2.162688(35)	& -2.203809(55)	& -2.245610(52)	 \\
 1.3  	& -1.851381(37)	& -1.926653(39)	& -2.003550(42)	& -2.082194(64)	& -2.163265(53)	& -2.204428(44)	& -2.246327(62)	 \\
$\infty$ 	& -1.85320(36)	& -1.92791(6)	& -2.00510(11)	& -2.08395(9)	& -2.16542(8)	& -2.20692(15)	& -2.24894(16)	\\
\hline
$ \chi^2$ & 0.9 	& 2.2 	& 2.8 	& 2.8 	& 1.9 	& 1.7 	& 0.9   \\
  \hline
  \end{tabular}
    \caption{Grand potential energy per site of the uniform phase at $U/t=8$ as a  function of $\tau$  
    for the $16\times16$ lattice, 
    extrapolated to the $\tau\to \infty$ limit with  the same 
    functional form  used in  Tab.~(\ref{tab1}).
    The last raw indicates the reduced $\chi^2$  of the  fit. The numbers in parenthesis
    represent error bars in the  last digits.}
   \label{tab16}
\end{table}

\begin{table}[h]
  \begin{tabular}{|l|l|l|l|l|l|}
    \hline
    $\tau t$ &  $\mu=1.75t$ & $\mu=1.8t$  & $\mu=1.85t$ & $\mu=1.9t$  & $\mu=1.95t$  \\
    \hline
 .8  & -2.282885(36)	& -2.325675(39)	& -2.369307(40)	& -2.413621(28)	& -2.458617(28) \\
 .9  & -2.284800(36)	& -2.327759(40)	& -2.371637(41)	& -2.416127(29)	& -2.461472(28) \\
 1.0 & -2.286276(38)	& -2.329368(44)	& -2.373345(42)	& -2.417994(30)	& -2.463533(30) \\
 1.1 & -2.287352(34)	& -2.330611(47)	& -2.374657(35)	& -2.419552(36)	& -2.465073(36) \\
 1.2 & -2.288204(38)	& -2.331600(50)	& -2.375645(27)	& -2.420577(46)	& -2.466332(46) \\
 1.3 & -2.288796(37)	& -2.332109(65)	& -2.376450(36)	& -2.421421(53)	& -2.467071(56) \\
$\infty$ & -2.29161(13)	& -2.33567(16)	& -2.37984(13)	& -2.42528(13)	& -2.47096(14) \\
\hline
$ \chi^2$ &  0.3 	& 0.1 	& 0.5 	& 1.8 	& 1.7  \\
  \hline
  \end{tabular}
    \caption{Grand potential energy per site of the uniform phase at $U/t=8$ as a  function of $\tau$  
    for the $16\times16$ lattice, 
    extrapolated to the $\tau\to \infty$ limit with  the same 
    functional form  used in  Tab.~(\ref{tab1}).
    The last raw indicates the reduced $\chi^2$  of the  fit. The numbers in parenthesis
    represent error bars in the  last digits.}
   \label{tab17}
\end{table}
\begin{table}[h]
  \begin{tabular}{|l|l|l|l|l|l|l|l|l|}
    \hline
    $\tau t$ &  $\mu=t$ & $\mu=1.1t$ & $\mu=1.2t$ & $\mu=1.3t$  & $\mu=1.35t$ & $\mu=1.4t$ & $\mu=1.45t$  \\
    \hline
 .93  	& -1.777339(29)	& -1.852837(30)	& -1.930114(30)	& -2.009134(28)	& -2.049657(28)	& -2.090429(32)	& -2.131616(31) \\
 1.06  	& -1.778247(31)	& -1.853795(31)	& -1.931060(31)	& -2.010177(29)	& -2.050684(29)	& -2.091531(31)	& -2.132778(31) \\
 1.20  	& -1.778682(32)	& -1.854269(32)	& -1.931698(31)	& -2.010944(29)	& -2.051549(28)	& -2.092334(32)	& -2.133689(32) \\
 1.33  	& -1.779112(33)	& -1.854715(32)	& -1.932187(32)	& -2.011430(30)	& -2.052058(30)	& -2.093002(34)	& -2.134354(34) \\
 1.46  	& -1.779315(32)	& -1.855027(33)	& -1.932542(34)	& -2.011702(32)	& -2.052372(32)	& -2.093365(36)	& -2.134806(36) \\
 1.60  	& -1.779553(35)	& -1.855194(36)	& -1.932808(37)	& -2.011983(35)	& -2.052772(36)	& -2.093730(40)	& -2.135282(42) \\
 1.73  	& -1.779702(39)	& -1.855359(42)	& -1.932967(47)	& -2.012251(50)	& -2.052841(46)	& -2.093871(57)	& -2.135510(34) \\
$\infty$ & -1.78020(7)	& -1.85619(12)	& -1.93380(7)	& -2.01293(7)	& -2.05378(7)	& -2.09497(8)	& -2.13684(11) \\
\hline
$\chi^2$ & 2.9 	& 2.1 	& 0.6 	& 1.2 	& 2.6 	& 1.5 	& 0.7 \\
  \hline
  \end{tabular}
    \caption{Grand potential energy per site of the uniform phase at $U/t=6$ as a  function of $\tau$  
    for the $16\times16$ lattice, 
    extrapolated to the $\tau\to \infty$ limit with  the same 
    functional form  used in  Tab.~(\ref{tab1}).
    The last raw indicates the reduced $\chi^2$  of the  fit. The numbers in parenthesis
    represent error bars in the  last digits.}
   \label{tab18}
\end{table}

\begin{table}[h]
  \begin{tabular}{|l|l|l|l|l|l|l|}
    \hline
    $\tau t$ &  $\mu=1.5t$ & $\mu=1.55t$  & $\mu=1.6t$ & $\mu=1.65t$  & $\mu=1.7t$ &  $\mu=1.75t$ \\
    \hline
 .93  	& -2.173287(33)	& -2.215428(33)	& -2.258169(34)	& -2.301278(34)	& -2.345671(34)	& -2.390396(34) \\
 1.06  	& -2.174613(36)	& -2.216892(36)	& -2.259813(36)	& -2.303334(36)	& -2.347832(35)	& -2.392994(36) \\
 1.20  	& -2.175495(36)	& -2.218034(33)	& -2.261165(36)	& -2.304980(36)	& -2.349560(36)	& -2.395029(36) \\
 1.33  	& -2.176184(38)	& -2.218740(37)	& -2.262002(37)	& -2.306008(38)	& -2.350806(38)	& -2.396350(38) \\
 1.46  	& -2.176776(39)	& -2.219441(40)	& -2.262729(41)	& -2.306779(42)	& -2.351770(43)	& -2.397492(45) \\
 1.60  	& -2.177208(34)	& -2.219884(36)	& -2.263252(53)	& -2.307496(41)	& -2.352489(42)	& -2.398408(45) \\
 1.73  	& -2.177593(41)	& -2.220202(41)	& -2.263606(47)	& -2.307823(47)	& -2.353054(49)	& -2.398972(44) \\
$\infty$ & -2.17929(13)	& -2.22170(8)	& -2.26530(9)	& -2.30979(8)	& -2.35547(8)	& -2.40169(8) \\
\hline
 $\chi^2$ & 0.3 	& 1.4 	& 0.5 	& 1.2 	& 0.4 	& 2.1  \\
  \hline
  \end{tabular}
    \caption{Grand potential energy per site of the uniform phase at $U/t=6$ as a  function of $\tau$  
    for the $16\times16$ lattice, 
    extrapolated to the $\tau\to \infty$ limit with  the same 
    functional form  used in  Tab.~(\ref{tab1}).
    The last raw indicates the reduced $\chi^2$  of the  fit. The numbers in parenthesis
    represent error bars in the  last digits.}
   \label{tab19}
\end{table}
\begin{table}[h]
  \begin{tabular}{|l|l|l|l|l|l|l|l|l|}
    \hline
    $\tau t$ &  $\mu=0.5t$ & $\mu=0.6t$ & $\mu=0.7t$ & $\mu=0.8t$  & $\mu=0.9t$ & $\mu=t$ & $\mu=1.1t$  \\
    \hline
 1.4  	& -1.521621(19)	& -1.595004(29)	& -1.670021(21)	& -1.746991(21)	& -1.826104(21)	& -1.907332(21)	& -1.990550(21) \\
 1.6  	& -1.522055(20)	& -1.595372(31)	& -1.670416(21)	& -1.747382(22)	& -1.826501(22)	& -1.907767(22)	& -1.991020(23) \\
 1.8  	& -1.522311(21)	& -1.595502(30)	& -1.670462(22)	& -1.747508(22)	& -1.826633(22)	& -1.907918(22)	& -1.991183(23) \\
 2.0  	& -1.522462(21)	& -1.595624(31)	& -1.670557(22)	& -1.747614(22)	& -1.826838(22)	& -1.908078(32)	& -1.991366(33) \\
 2.2  	& -1.522571(21)	& -1.595688(31)	& -1.670622(22)	& -1.747696(22)	& -1.826888(23)	& -1.908131(24)	& -1.991522(25) \\
 2.4  	& -1.522634(21)	& -1.595729(31)	& -1.670708(22)	& -1.747752(23)	& -1.826964(24)	& -1.908257(25)	& -1.991593(27) \\
 2.6  	& -1.522710(22)	& -1.595852(32)	& -1.670908(33)	& -1.747829(24)	& -1.827068(26)	& -1.908379(29)	& -1.991737(23) \\
$\infty$ & -1.52288(4)	& -1.59609(11)	& -1.67112(8)	& -1.74808(8)	& -1.82736(8)	& -1.90876(9)	& -1.99223(8) \\
\hline
 $\chi^2$& 0.2 	& 0.9 	& 0.1 	& 0.2 	& 2.3 	& 1.6 	& 0.8  \\
  \hline
  \end{tabular}
    \caption{Grand potential energy per site of the uniform phase at $U/t=4$ as a  function of $\tau$  
    for the $16\times16$ lattice, 
    extrapolated to the $\tau\to \infty$ limit with  the same 
    functional form  used in  Tab.~(\ref{tab1}).
    The last raw indicates the reduced $\chi^2$  of the  fit. The numbers in parenthesis
    represent error bars in the  last digits.}
   \label{tab18}
\end{table}

\begin{table}[h]
  \begin{tabular}{|l|l|l|l|l|}
    \hline
    $\tau t$ &  $\mu=1.2t$ & $\mu=1.25t$  & $\mu=1.3t$ & $\mu=1.4t$ \\
    \hline
 1.4  	& -2.075581(21)	& -2.117264(30)	& -2.162617(22)	& -2.252046(22) \\
 1.6  	& -2.076076(22)	& -2.119177(32)	& -2.163309(23)	& -2.253041(22) \\
 1.8  	& -2.076356(23)	& -2.119344(32)	& -2.163678(22)	& -2.253764(23) \\
 2.0  	& -2.076590(33)	& -2.119981(33)	& -2.164039(33)	& -2.254322(34) \\
 2.2  	& -2.076723(25)	& -2.120221(35)	& -2.164250(25)	& -2.254835(26) \\
 2.4  	& -2.076880(28)	& -2.120297(41)	& -2.164509(30)	& -2.255166(36) \\
 2.6  	& -2.077040(37)	& -2.120419(44)	& -2.164670(45)	& -2.255440(65) \\
 $\infty$ & -2.07736(6)	& -2.12071(15)	& -2.16550(11)	& -2.25659(7) \\
\hline
$\chi^2$ & 1.9 	& 0.6 	& 0.8 	& 1.3  \\
  \hline
  \end{tabular}
    \caption{Grand potential energy per site of the uniform phase at $U/t=4$ as a  function of $\tau$  
    for the $16\times16$ lattice, 
    extrapolated to the $\tau\to \infty$ limit with  the same 
    functional form  used in  Tab.~(\ref{tab1}).
    The last raw indicates the reduced $\chi^2$  of the  fit. The numbers in parenthesis
    represent error bars in the  last digits.}
   \label{tab19}
\end{table}
\begin{table}[h]
  \begin{tabular}{|l|l|l|l|l|l|l|l|l|l|}
    \hline
 $\tau t$ & $\mu=1.4t$ & $\mu=1.45t$ & $\mu=1.5t$ & $\mu=1.55t$  & $\mu=1.6t$ & $\mu=1.65t$ & $\mu=1.7t$ & $\mu=1.75t$   \\
    \hline
  .8     & -2.000359(33) & -2.039516(33) & -2.078710(34) & -2.118534(34) & -2.158590(34) & -2.199322(34) & -2.240738(33) & -2.282806(34)   \\
  .9     & -2.001536(34) & -2.040696(33) & -2.080192(34) & -2.120028(35) & -2.160159(34) & -2.201043(34) & -2.242484(34) & -2.284816(34)   \\
 1.0    & -2.002488(36) & -2.041695(35) & -2.081147(36) & -2.121035(37) & -2.161442(36) & -2.202324(35) & -2.243935(36) & -2.286245(36)    \\
 1.1    & -2.003178(38) & -2.042342(39) & -2.081877(38) & -2.121849(39) & -2.162265(38) & -2.203350(38) & -2.244920(39) & -2.287398(40)    \\
 1.2    & -2.003514(44) & -2.042753(43) & -2.082332(42) & -2.122375(43) & -2.162821(43) & -2.203934(44) & -2.245667(45) & -2.288183(28)    \\
 1.3    & -2.004004(39) & -2.043117(54) & -2.082678(55) & -2.122830(55) & -2.163391(57) & -2.204507(60) & -2.246324(63) & -2.288870(41)    \\
 $\infty$     & -2.00559(13)  & -2.04469(14)  & -2.08408(14)  & -2.12463(15)  & -2.16543(15)  & -2.20684(15)  & -2.24900(15)  & -2.29164(12)   \\
  \hline  
 $\chi^2$     & 3.1   & 2.3   & 0.3   & 0.5   & 2.5   & 0.25   & 1.3   & 0.8   \\
    \hline
  \end{tabular}
    \caption{Grand potential energy per site of the uniform phase as a  function of $\tau$ 
    limit for the $12\times 12$ lattice at $U/t=8$, extrapolated to the $\tau\to \infty$ limit with  the same
     function used in  Tab.~(\ref{tab1}). 
     The last raw indicates the reduced $\chi^2$  of the  fit. 
    The numbers in parenthesis represent error bars in the  last digits.}
 \label{tab12a}
\end{table}

\begin{table}[h]
  \begin{tabular}{|l|l|l|l|l|}
    \hline
    $\tau t$ & $\mu=1.8t$  & $\mu=1.85t$ & $\mu=1.9t$  & $\mu=1.95t$  \\
    \hline
  .8    & -2.325598(33) & -2.369161(33) & -2.413485(34) & -2.458440(33)  \\
  .9    & -2.327695(34) & -2.371545(35) & -2.416002(34) & -2.461283(33)  \\
 1.0    & -2.329395(37) & -2.373231(36) & -2.417883(36) & -2.463349(36)  \\
 1.1    & -2.330522(40) & -2.374608(41) & -2.419328(41) & -2.464898(40)  \\
 1.2    & -2.331498(48) & -2.375556(49) & -2.420510(50) & -2.466220(48)  \\
 1.3    & -2.332058(48) & -2.376196(63) & -2.421350(50) & -2.467025(57)  \\
 $\infty$  & -2.33512(14)  & -2.37945(16)  & -2.42525(14)  & -2.47110(15)  \\
  \hline  
 $\chi^2$   & 2.5   & 1.4   & 0.6   & 1.8 \\
    \hline
  \end{tabular}
    \caption{Grand potential energy per site of the uniform phase as a  function of $\tau$ 
    limit for the $12\times 12$ lattice at $U/t=8$
 extrapolated to the $\tau\to \infty$ limit with  the same
     function used in  Tab.~(\ref{tab1}).
     The last raw indicates the reduced $\chi^2$  of the  fit. 
    The numbers in parenthesis represent error bars in the  last digits.}
 \label{tab12b}
\end{table}

\FloatBarrier

\section{ Stripe wave function  parameterization}
The variational parameters of the 
mean field Hamiltonian  in this case depend explicitly on $r_x$, satisfying 
periodicity or antiperiodicity properties once $r_x$ is changed to $r_x+W$ as described in the main text.
$H_{MF}(\bm{\alpha})$ is defined therefore by:
\begin{eqnarray}
H_{MF}(\bm{\alpha}) &=& \sum\limits_{r_x,r_y,\sigma} \left[-t_x(r_x) c^\dag_{r_x+1,r_y,\sigma} c_{r_x,r_y,\sigma}-t_y(r_x) c^\dag_{r_x,r_y+1,\sigma} c_{r_x,r_y,\sigma} + {\rm h.c.} \right] \nonumber \\
&+& \sum\limits_{r_x,r_y} \Delta_{AF}(r_x) (-1)^{r_x+r_y} (n_{\uparrow,r_x,r_y}-n_{\downarrow,r_x,r_y}) -\mu(r_x) n_{r_x,r_y} 
\end{eqnarray}
where $n_{\uparrow,r_x,r_y}=c^\dag_{\uparrow,r_x,r_y} c_{\uparrow,r_x,r_y}$,
$n_{\downarrow,r_x,r_y}=c^\dag_{\downarrow,r_x,r_y} c_{\downarrow,r_x,r_y}$,  
$n_{r_x,r_y}=n_{\uparrow,r_x,r_y}+n_{\downarrow,r_x,r_y}$, and $r_x,r_y$ are integer indices 
describing the lattice. $t_x$,  $t_y$ and $\mu$ periodic in $r_x$, while 
$\Delta_{AF}$ antiperiodic.
Also in this case the optimization of $r_x-$periodic $t^\prime$ and 
BCS pairing $\Delta_{BCS}$ were tried, by adding the following terms to $H_{MF}(\bm{\alpha})$:
\begin{eqnarray}
\delta H_{MF}&=&\sum\limits_{r_x,r_y,\sigma} \left[-t^\prime_x(r_x) c^\dag_{r_x+1,r_y+1,\sigma} c_{r_x,r_y,\sigma}-t^\prime_y(r_x) c^\dag_{r_x+1,r_y-1,\sigma} c_{r_x,r_y,\sigma} + {\rm h.c.} \right] \\
&+& \sum\limits_{r_x,r_y} \left[ \Delta_{BCS}^x(r_x) ( c^\dag_{r_x+1,r_y,\uparrow} c^\dag_{r_x,r_y,\downarrow} + c^\dag_{r_x,r_y,\uparrow} c^\dag_{r_x+1,r_y,\downarrow} ) 
-\Delta_{BCS}^y(r_x)  ( c^\dag_{r_x,r_y+1,\uparrow} c^\dag_{r_x,r_y,\downarrow}
+c^\dag_{r_x,r_y,\uparrow} c^\dag_{r_x,r_y+1,\downarrow}) 
+ {\rm h.c.} \right] \nonumber \\
\end{eqnarray}
with further $4 W$ variational parameters for $H_{MF}(\bm{\alpha})$.  However both $t^\prime$ and 
pairing  were vanishingly small and very difficult to stabilize, providing 
no sizable improvement to the energy.
\begin{table}[h]
  \begin{tabular}{|l|l|l|l|l|l|l|l|}
    \hline
    $\tau t$ & $N=114$  & $N=116$ & $N=118$  & $N=120$  & $N=122$ & $N=124$ & $N=126$  \\
    \hline
 .8  	& -0.890519(28)	& -0.869528(28)	& -0.847996(27)	& -0.826324(27)	& -0.802049(27)	& -0.778540(27)	& -0.753773(26) \\
 .9  	& -0.892160(28)	& -0.871321(28)	& -0.849528(28)	& -0.827634(28)	& -0.803228(28)	& -0.779806(28)	& -0.755058(28) \\
 1.0  	& -0.893276(29)	& -0.872410(29)	& -0.850662(30)	& -0.828763(30)	& -0.804204(30)	& -0.780403(30)	& -0.756032(29) \\
 1.1  	& -0.894136(31)	& -0.873049(31)	& -0.851298(31)	& -0.829488(32)	& -0.804938(32)	& -0.780895(32)	& -0.756670(32) \\
 1.2  	& -0.894604(34)	& -0.873654(36)	& -0.851896(36)	& -0.830080(36)	& -0.805432(37)	& -0.781291(37)	& -0.757158(36) \\
 1.3  	& -0.895200(45)	& -0.874050(47)	& -0.852185(47)	& -0.830573(44)	& -0.806029(48)	& -0.781732(47)	& -0.757696(44) \\
$\infty$ & -0.89695(12)	& -0.87529(12)	& -0.85371(12)	& -0.83265(12)	& -0.80804(12)	& -0.78353(22)	& -0.75934(12) \\
  \hline  
$ \chi^2$ & 3.5 	& 3.3 	& 2.8 	& 1.8 	& 2.3 	& 1.3 	& 2.7  \\
    \hline
  \end{tabular}
    \caption{Energy per site of the $W=6$ length stripe at $U/t=8$ as a  function of $\tau$  for the $12\times 12$ lattice
 extrapolated to the $\tau\to \infty$ limit with  the same
     function used in  Tab.~(\ref{tab1}). The last raw indicates the reduced $\chi^2$  of the  fit. 
    The numbers in parenthesis represent error bars in the  last digits.}
\end{table}

\begin{table}[h]
  \begin{tabular}{|l|l|l|l|l|l|}
    \hline
    $\tau t$ & $N=164$  & $N=166$ & $N=168$  & $N=170$  & $N=172$  \\
    \hline
.8     & -0.819233(29) & -0.802616(29) & -0.786014(28) & -0.768011(28) & -0.749930(28) \\
.9     & -0.820719(29) & -0.804046(29) & -0.787349(29) & -0.769293(29) & -0.751120(28) \\
1.0    & -0.821920(32) & -0.805098(31) & -0.788383(31) & -0.770310(31) & -0.752117(31) \\
1.1    & -0.822778(33) & -0.805944(33) & -0.789093(33) & -0.771040(33) & -0.752790(33) \\
1.2    & -0.823281(39) & -0.806436(38) & -0.789618(38) & -0.771534(39) & -0.753403(38) \\
1.3    & -0.823861(55) & -0.806936(52) & -0.790104(35) & -0.772046(35) & -0.753826(50) \\
$\infty$    & -0.82592(13)  & -0.80887(13)  & -0.79195(11)  & -0.77395(11)  & -0.75585(13) \\
  \hline  
$\chi^2$    & 3.0   & 1.3   & 0.8   & 1.2   & 0.8 \\
    \hline
  \end{tabular}
    \caption{Energy per site of the $W=7$ length stripe at $U/t=8$ as a  function of $\tau$  for the $14\times 14$ lattice
 extrapolated to the $\tau\to \infty$ limit with  the same
     function used in  Tab.~(\ref{tab1}). The last raw indicates the reduced $\chi^2$  of the  fit. 
    The numbers in parenthesis represent error bars in the  last digits.}
\end{table}

\begin{table}[h]
  \begin{tabular}{|l|l|l|l|l|l|l|}
    \hline
    $\tau t$ & $N=212$  & $N=216$ & $N=218$  & $N=220$  & $N=222$ & $N=224$  \\
    \hline
 .8  	& -0.832104(28)	& -0.806497(28)	& -0.793530(28)	& -0.780653(27)	& -0.767527(35)	& -0.754612(35)	 \\
 .9  	& -0.833601(29)	& -0.807988(28)	& -0.794942(28)	& -0.781919(28)	& -0.768870(28)	& -0.755811(36)	 \\
 1.0  	& -0.834915(30)	& -0.809198(31)	& -0.796119(31)	& -0.782952(31)	& -0.769884(38)	& -0.756767(40)	 \\
 1.1  	& -0.835729(34)	& -0.809983(34)	& -0.796882(34)	& -0.783771(33)	& -0.770591(41)	& -0.757333(35)	 \\
 1.2  	& -0.836307(40)	& -0.810497(39)	& -0.797447(40)	& -0.784324(41)	& -0.771078(38)	& -0.757880(39)	 \\
 1.3  	& -0.836988(61)	& -0.811081(60)	& -0.798044(60)	& -0.784684(54)	& -0.771511(51)	& -0.758326(37)	 \\
$ \infty$ & -0.83919(14)	& -0.81300(13)	& -0.80009(13)	& -0.78682(13)	& -0.77320(13)	& -0.76004(13) \\
    \hline
 $\chi^2$ & 5.8 	& 3.2 	& 2.5 	& 1.8 	& 0.4 	& 1.9 	 \\
  \hline  
  \end{tabular}
    \caption{Energy per site of the $W=8$ length stripe at $U/t=8$ as a  function of $\tau$  for the $16\times 16$ lattice
 extrapolated to the $\tau\to \infty$ limit with  the same
     function used in  Tab.~(\ref{tab1}). 
     The last raw indicates the reduced $\chi^2$  of the  fit. 
    The numbers in parenthesis represent error bars in the  last digits.}
\end{table}

\begin{table}[h]
  \begin{tabular}{|l|l|l|l|l|l|l|l|}
    \hline
    $\tau t$ & $N=226$ & $N=228$ & $N=230$  & $N=232$ & $N=234$ & $N=236$  & $N=240$   \\
    \hline
 .8  	& -0.740516(34)	& -0.726463(27)	& -0.712162(27)	& -0.697903(26)	& -0.683110(26)	& -0.668374(26)	& -0.637611(26) \\
 .9  	& -0.741674(36)	& -0.727599(29)	& -0.713277(28)	& -0.698880(28)	& -0.684164(27)	& -0.669375(28)	& -0.638553(28) \\
 1.0  	& -0.742488(39)	& -0.728462(31)	& -0.714098(30)	& -0.699703(31)	& -0.685025(30)	& -0.670182(30)	& -0.639358(31) \\
 1.1  	& -0.743116(41)	& -0.729033(32)	& -0.714736(32)	& -0.700350(33)	& -0.685603(32)	& -0.670776(32)	& -0.640024(33) \\
 1.2  	& -0.743756(54)	& -0.729633(38)	& -0.715294(37)	& -0.700903(39)	& -0.686195(38)	& -0.671300(38)	& -0.640505(38) \\
 1.3  	& -0.744120(47)	& -0.730092(52)	& -0.715803(49)	& -0.701403(37)	& -0.686582(48)	& -0.671742(52)	& -0.641135(50) \\
$ \infty$ & -0.74596(14)	& -0.73184(13)	& -0.71764(12)	& -0.70347(11)	& -0.68841(12)	& -0.67352(13)	& -0.64314(12) \\
    \hline
 $\chi^2$ 	& 1.8 	& 2.9 	& 2.4 	& 1.1 	& 1.4 	& 0.7 	& 3.5  \\
  \hline  
  \end{tabular}
    \caption{Energy per site of the $W=8$ length stripe at $U/t=8$ as a  function of $\tau$  for the $16\times 16$ lattice 
     extrapolated to the $\tau\to \infty$ limit with  the same
         function used in  Tab.~(\ref{tab1}).
	 The last raw indicates the reduced $\chi^2$  of the  fit. 
    The numbers in parenthesis represent error bars in the  last digits.}
\end{table}
\begin{table}[h]
  \begin{tabular}{|l|l|l|l|l|l|l|}
    \hline
    $\tau t$ & $N=192$ & $N=194$ & $N=196$  & $N=198$ & $N=200$ & $N=202$   \\
    \hline
 .7  	& -0.727826(40)	& -0.710866(36)	& -0.694006(36)	& -0.676730(25)	& -0.659492(25)	& -0.641348(25) \\
 .8  	& -0.729478(44)	& -0.712487(38)	& -0.695612(39)	& -0.678313(27)	& -0.661011(27)	& -0.642843(27) \\
 .9  	& -0.730601(45)	& -0.713653(40)	& -0.696686(40)	& -0.679361(28)	& -0.662029(28)	& -0.643818(28) \\
 1.0  	& -0.731403(48)	& -0.714406(43)	& -0.697420(43)	& -0.680076(31)	& -0.662629(31)	& -0.644429(29) \\
 1.1  	& -0.732250(52)	& -0.715196(47)	& -0.698188(47)	& -0.680846(33)	& -0.663374(33)	& -0.645146(33) \\
 1.2  	& -0.732705(61)	& -0.715720(51)	& -0.698688(39)	& -0.681350(37)	& -0.663939(37)	& -0.645735(37) \\
 1.3  	& -0.733135(77)	& -0.716203(68)	& -0.699243(47)	& -0.681840(44)	& -0.664389(46)	& -0.646196(44) \\
$\infty$	& -0.73481(11)	& -0.71817(17)	& -0.70182(27)	& -0.68432(22)	& -0.66721(23)	& -0.64905(22) \\
\hline 
 $\chi^2$	& 2.5 	& 2.1 	& 1.4 	& 2.0 	& 2.0 	& 1.1  \\
  \hline  
  \end{tabular}
    \caption{Energy per site of the $W=9$ length stripe at $U/t=8$ as a  function of $\tau$  for the $18\times 12$ lattice 
     extrapolated to the $\tau\to \infty$ limit with  the same
         function used in  Tab.~(\ref{tab1}).
	 The last raw indicates the reduced $\chi^2$  of the  fit. 
    The numbers in parenthesis represent error bars in the  last digits.}
\end{table}

\begin{table}[h]
  \begin{tabular}{|l|l|l|l|l|}
    \hline
    $\tau t$ & $N=288$ & $N=292$ & $N=296$  & $N=300$   \\
    \hline
 .7  	& -0.664048(41)	& -0.640305(21)	& -0.616020(21)	& -0.590547(21) \\
 .8  	& -0.665373(45)	& -0.641484(23)	& -0.617236(23)	& -0.591674(24) \\
 .9  	& -0.666597(47)	& -0.642713(23)	& -0.618582(24)	& -0.593141(24) \\
 1.0  	& -0.667202(38)	& -0.643460(25)	& -0.619281(25)	& -0.593909(26) \\
 1.1  	& -0.667958(32)	& -0.643997(29)	& -0.619727(27)	& -0.594405(28) \\
 1.2  	& -0.668511(34)	& -0.644520(33)	& -0.620175(33)	& -0.594796(32) \\
 1.3  	& -0.668864(43)	& -0.644865(43)	& -0.620594(39)	& -0.595044(39) \\
$\infty$ & -0.67028(7)	& -0.64665(19)	& -0.62315(38)	& -0.59606(18) \\
\hline
 $\chi^2$ & 0.2 	& 1.2 	& 0.1 	& 0.2  \\
  \hline  
  \end{tabular}
    \caption{Energy per site of the $W=13$ length stripe at $U/t=8$ as a  function of $\tau$  for the $26\times 12$ lattice 
     extrapolated to the $\tau\to \infty$ limit with  the same
         function used in  Tab.~(\ref{tab1}).
	 The last raw indicates the reduced $\chi^2$  of the  fit. 
    The numbers in parenthesis represent error bars in the  last digits.}
\end{table}

\FloatBarrier

\begin{table}[h]
  \begin{tabular}{|l|l|l|l|l|l|}
    \hline
    $\tau t$ &  $12\times 12$  & $14\times14$ & $16\times 16$ & $18\times 12$ & $26\times 12$   \\
 \hline
 .8  	& -0.826324(27)	 & -0.786044(29)	& -0.754606(18)	& -0.729477(44)	& -0.665351(32)  \\
 .9  	& -0.827634(28)	 & -0.787398(31)	& -0.755791(19)	& -0.730598(45)	& -0.666573(33)  \\
 1.0  	& -0.828763(30)	 & -0.788455(33)	& -0.756692(21)	& -0.731405(48)	& -0.667280(36)  \\
 1.1  	& -0.829488(32)	 & -0.789069(35)	& -0.757351(22)	& -0.732245(53)	& -0.667995(40)  \\
 1.2  	& -0.830080(36)	 & -0.789641(40)	& -0.757845(26)	& -0.732692(61)	& -0.668456(47)  \\
 1.3  	& -0.830573(44)	 & -0.790096(51)	& -0.758339(33)	& -0.733106(77)	& -0.668863(61)  \\
 $\infty$ & -0.83265(12)	 & -0.79177(13)	& -0.76007(8)	& -0.73477(11)	& -0.67056(9)  \\
\hline				
 $\chi^2$ & 1.8 	 & 2.0 	& 1.9 	& 2.3 	& 2.7   \\
  \hline  
  \end{tabular}
    \caption{Energy per site of the $W={1\over \delta}$ length stripe at $U/t=8$ as a  function of $\tau$  for $W=6,7,8,9,13$ from left column to right column, respectively.
These values are extrapolated to the $\tau\to \infty$ limit with  the same
         function used in  Tab.~(\ref{tab1}).
	 The last raw indicates the reduced $\chi^2$  of the  fit. 
    The numbers in parenthesis represent error bars in the  last digits.}
\end{table}
\begin{table}[h]
  \begin{tabular}{|l|l|l|l|l|}
    \hline
    $\tau t$ &  $12\times 12$  & $14\times14$ & $16\times 16$ & $18\times 12$ \\
 \hline
 1.06  	& -0.925307(24)	 & -0.890645(25)	& -0.863055(30)	 & -0.841025(32) \\
 1.20  	& -0.925774(25)	 & -0.891330(26)	& -0.863805(33)	 & -0.841656(32) \\
 1.33  	& -0.926254(26)	 & -0.891887(28)	& -0.864303(35)	 & -0.842194(38) \\
 1.46  	& -0.926620(28)	 & -0.892257(29)	& -0.864598(37)	 & -0.842488(40) \\
 1.60  	& -0.926951(33)	 & -0.892557(33)	& -0.864869(42)	 & -0.842804(47) \\
 1.73  	& -0.927107(43)	 & -0.892873(44)	& -0.865129(42)	 & -0.843116(44) \\
$\infty$ & -0.92837(11)	 & -0.89396(11)	& -0.86587(8)	 & -0.84418(13) \\
\hline			
$\chi^2$ & 1.4 	 & 1.1 	& 0.7 	 & 1.4  \\
  \hline  
  \end{tabular}
    \caption{Energy per site of the $W={1\over \delta}$ length stripe at $U/t=6$ as a  function of $\tau$  for $W=6,7,8,9$ from left column to right column, respectively.
These values are extrapolated to the $\tau\to \infty$ limit with  the same
         function used in  Tab.~(\ref{tab1}).
	 The last raw indicates the reduced $\chi^2$  of the  fit. 
    The numbers in parenthesis represent error bars in the  last digits.}
\end{table}

\begin{table}[h]
  \begin{tabular}{|l|l|l|l|l|}
    \hline
    $\tau t$ & $14\times14$ & $16\times 16$ & $18\times 12$ & $26\times 16$ \\
 \hline
 1.4  	& -1.046762(12)	& -1.025782(22)	& -1.008801(24)	& -0.964547(13) \\
 1.6  	& -1.047262(13)	& -1.026152(22)	& -1.009235(25)	& -0.964839(13) \\
 1.8  	& -1.047511(13)	& -1.026359(23)	& -1.009477(24)	& -0.965136(27) \\
 2.0  	& -1.047736(13)	& -1.026621(23)	& -1.009757(26)	& -0.965349(14) \\
 2.2  	& -1.047902(14)	& -1.026632(25)	& -1.009871(27)	& -0.965490(16) \\
 2.4  	& -1.048022(16)	& -1.026754(28)	& -1.009891(31)	& -0.965600(19) \\
 2.6  	& -1.048105(19)	& -1.026816(37)	& -1.010068(38)	& -0.965685(25) \\
$\infty$ & -1.04857(5)	& -1.02706(6)	& -1.01035(6)	& -0.96614(4) \\
\hline			
 $\chi^2$ & 0.5 	& 3.1 	& 2.7 	& 1.3  \\
  \hline  
  \end{tabular}
    \caption{Energy per site of the $W={1\over \delta}$ length stripe at $U/t=4$ as a  function of $\tau$  for $W=7,8,9,13$ from left column to right column, respectively.
These values are extrapolated to the $\tau\to \infty$ limit with  the same
         function used in  Tab.~(\ref{tab1}).
	 The last raw indicates the reduced $\chi^2$  of the  fit. 
    The numbers in parenthesis represent error bars in the  last digits.}
\end{table}

\bibliographystyle{apsrev4-1}
\bibliography{references.bib}